\documentclass[prd,tightenlines,showpacs,letterpaper, preprintnumbers,nofootinbib,floatfix]{revtex4}

\usepackage{amsmath,amssymb,amsfonts,graphicx,pifont,dcolumn}
\usepackage{epstopdf}
\usepackage{color}
\usepackage{cancel}
\usepackage{ulem}
\usepackage{xfrac}
\RequirePackage{slashed}

\begin{document}

\title{Signatures of excited monopolium}

\author{Huner Fanchiotti}  \author{Carlos A.~Garc\'\i a Canal}
\affiliation{ IFLP(CONICET) and Department of Physics, University of La Plata,  C.C. 67 1900, La Plata, Argentina}

\author{Marco~Traini} 

\affiliation{Dipartimento di Fisica, Universit\`a degli Studi di Trento, Via Sommarive 14, I-38123 Povo (Trento), Italy}

\author{ Vicente Vento }

\affiliation{Departamento de F\'{\i}sica Te\'orica and IFIC, Universidad de Valencia - CSIC, E-46100 Burjassot (Valencia), Spain.}

\date{\today}

\begin{abstract}
We study electromagnetic properties of particles with magnetic moment and no charge using their behavior when traversing coils and solenoids. These particles via the Faraday-Lenz law  create a current whose energy we calculate. We analyze both the case of very long lived, almost stable, particles and those with a finite lifetime. We use this development to study the behavior of monopolium a monopole-antimonopole bound state in its excited states. 

\end{abstract}


\maketitle

\section{Introduction}

Monopoles and their detection has been a matter of much research since Dirac discovered his now famous quantization condition~\cite{Dirac:1931qs,Dirac:1948um},

\begin{equation}
e g =\frac{N}{2} \;\;\; N= 1,2,\ldots,
\end{equation}
where $e$ is the electron charge, $g$ the monopole magnetic charge and we have used natural units $\hbar=c=4\pi \varepsilon_0=1$. 

In Dirac's formulation, monopoles are assumed to exist as point-like particles and quantum mechanical consistency conditions lead to establish the magnitude of their magnetic charge. Monopole physics took a dramatic turn when 't Hooft~\cite{tHooft:1974kcl} and Polyakov~\cite{Polyakov:1974ek} independently discovered that the SO(3) Georgi-Glashow model~\cite{Georgi:1972cj} inevitably contains monopole solutions. These topological monopoles are impossible to create in particle collisions because of their huge GUT scale mass~\cite{Preskill:1984gd} or in other models with lower mass because of their complicated multi-particle structure~\cite{Drukier:1981fq}. 

In Dirac's formulation, since the magnetic charge is conserved, monopoles  should be produced predominantly in monopole-antimonopole pairs. 
Inspired by the old idea of Dirac and Zeldovich~\cite{Dirac:1931qs,Zeldovich:1978wj} that monopoles are not seen free because they are confined by their strong magnetic forces, we have studied a monopole-antimonopole bound state that we have called monopolium~\cite{Vento:2007vy}. This state is the strongly coupled dual analog of positronium and decays into photons \cite{Epele:2012jn}.

The discovery of monopole and dipole solutions in Kaluza Klein theories \cite{Kaluza:1921tu,Klein:1926tv} made these theories very exciting
from a theoretical point of view \cite{Gross:1983hb}. In particular the dipole solution, which is classically stable and therefore very long lived, forms a very
interesting state which we have also called monopolium in analogy with its decaying cousin in  gauge
theories. The Kaluza Klein monopolium is extremely massive with a mass of the order of the Planck mass and therefore impossible to produce in laboratories. However, there might be clouds of them in the cosmos which might enter our detectors~\cite{Vento:2020vsq}.

Much experimental research has been carried out into the search for monopoles \cite{Cabrera:1982gz,Milton:2006cp,MoEDAL:2014ttp,Acharya:2014nyr,Patrizii:2015uea}. However, our interest here lies in the detection of monopolium, who is chargeless but which can manifest a magnetic moment in the presence of magnetic fields or in excited deformed states \cite{Vento:2020vsq,Vento:2019auh}.  Our aim here is to study the behavior of monopolium in the presence of coils and solenoids to determine interesting properties and energy regimes of excited monopolia which may serve as a guide to more sophisticated experiments given the progress of electromagnetic technologies at present.

In what follows we are going to discuss methods to study magnetic moments and how they might be used in the study of particle properties. In Section \ref{faraday}  we describe the effects of a particle with a magnetic dipole moment when traversing conducting  coils. In section \ref{Sdetector} we determine the observables characterizing  particles with magnetic moment and no electromagnetic charges. In section \ref{monopolium}  we describe the behavior of monopolium when traversing coils and solenoids. In section \ref{cosmic} we apply the unveiled  properties of monopolium to describe possible signatures of cosmic monopolium. We end in section \ref{conclusion} by collecting the most significant results of our investigation.

 \section{Effects of a particle with magnetic dipole moment  passing through a conducting coil}
\label{faraday}
Our subject of study is a particle which has a permanent or an induced magnetic dipole moment $\overrightarrow{\cal{M}}$. We call this particle dipole for short.  Assuming the size of the particle small compared with the distance at which we are measuring the magnetic field its dipole moment will produce a magnetic field \cite{Vento:2019auh}

\begin{equation}
\overrightarrow{B}_d (\vec{r})= \frac{ 3(\overrightarrow{\cal{M}}\cdot \vec{r}) \,\vec{r} - \overrightarrow{\cal{M}}\, r^2}{r^5},
\label{Bfield}
\end{equation}
where $r=|\vec{r}|= \sqrt{(x-x_0)^2 +(y-y_0)^2 +(z-z_0)^2}$, $(x,y,z)$ is any point in space and $(x_0, y_0, z_0)$ is the position of the dipole considered point-like for this calculation.

\begin{figure}[htb]
\begin{center}
\includegraphics[scale= 0.35]{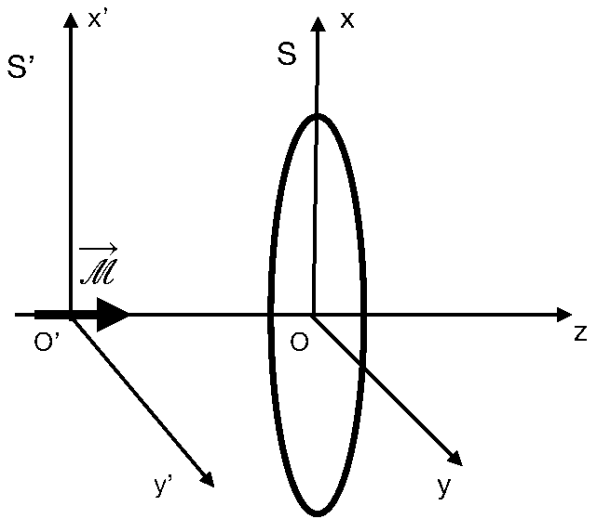} 
\end{center}
\caption{Our set up: a particle traveling with constant velocity  towards a conducting coil with its magnetic dipole moment perpendicular to the plane of the coil. $S^\prime$ is the reference system associated to the dipole and $S$ the one associated to the coil.}
\label{coil}
\end{figure}

Let the dipole travel with a constant velocity towards a circular conducting coil of radius $R$ with its magnetic moment perpendicular to the plane of the coil as shown in Fig. \ref{coil}. Let $S^\prime$ be the reference system associated to the dipole and $S$ the reference system associated to the coil. We choose the coil to be located in  the $z=0$ plane, thus at time $t$ the dipole will be located at $z= - v t$.
The magnetic field felt by any point of space $\vec{r^\prime} = (x^\prime, y^\prime, z^\prime)$ when the dipole is  $\overrightarrow{\cal{M}} = {\cal M}\, \hat{k^\prime}$ is

\begin{equation}
\overrightarrow{B^\prime}_d (\vec{r^\prime})= \frac{3 \mathcal{M} (\hat{k^\prime} \cdot \vec{r^\prime}) \,\vec{r^\prime} -  {\mathcal M} \, \hat{k^\prime} \, r^{\prime 2}}{r^{\prime 5}},
\label{Bfield0}
\end{equation}
where  $r^\prime  = |\vec{r^\prime}|= \sqrt{x^{\prime 2} + y^{\prime 2} +z^{\prime 2}}$ and $\hat{k^\prime}$ is the unit vector in the $z$ direction. We need the magnetic field at the coil. The Lorentz transformation from reference system $S^\prime$ to the reference system $S$  is given by,

\begin{align*}
x^\prime = & x, \\
y^\prime = & y,\\
z^\prime = & \gamma (z -v t), \\
t^\prime =& \gamma(t - z),\\
\end{align*}
where $\gamma = 1/\sqrt{1- v^2}$. Since we have no electric field present the magnetic field changes \cite{Apyan:2007wq}

\begin{align}
B^\prime_x &  = \gamma B_x,\\
B^\prime_y & = \gamma B_y,\\
B^\prime_z & = B_z. 
\end{align}

With this geometry the flux through the coil is

\begin{equation}
\Phi(t) = \int^R_0 d \rho \int^{2 \pi}_0 d \varphi \,\overrightarrow{B}_d (x , y, - \gamma v t) \cdot \hat{k},
\label{flux}
\end{equation}
where $\rho$ and $\varphi$ are the polar coordinates in the $z=0$ plane of the coil and $R$ is the radius of the coil. This calculation can be performed analytically and leads to

\begin{equation}
\Phi(t) = -\frac{2 {\cal M} \pi R^2}{((\gamma v t)^2 +R^2)^{3/2}}.
\label{flux1}
\end{equation}
\begin{figure}[htb]
\begin{center}
\includegraphics[scale= 0.9]{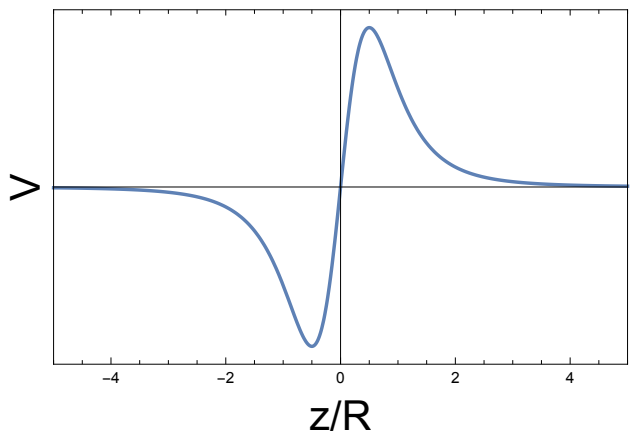} 
\end{center}
\caption{Functional form of the voltage generated by a dipole with a magnetic moment  passing a conducting coil at uniform speed.}
\label{emffig}
\end{figure}

Applying the Faraday-Lenz law the induced Electro Motive Force (EMF) generated by the magnetic field becomes

\begin{equation}
V(\gamma\, v\,t,R) =  \frac{ 6 {\cal M} \pi R^2 (\gamma v )^2 t}{((\gamma v t)^2 +R^2)^{5/2}}.
\label{emf}
\end{equation}
This equation can be written using $z(t) =  \gamma v t$, the area of the coil $A=\pi R^2$ and the strength parameter $\chi = 6 {\mathcal M} \gamma v$ as

\begin{equation}
V(z, A,\chi) =  \frac{\chi A z }{(z^2 + \frac{A}{\pi})^{\sfrac{5}{2}}}.
\label{emf1}
\end{equation}
This function is presented in Fig.\ref{emffig} for fixed $A$ and $\chi$.

We have assumed in the previous calculation that the dipole passes through the axis of the coil. Ref. \cite{Apyan:2007wq} shows the result when the dipole does not pass through the axis, which is non analytic. Off axis the shape of the pulse is asymmetric and the time span shorter. For the purposes of the estimates which will be calculated in this presentation Eq.(\ref{emf1}) will be sufficient, with the advantage of being analytic.

A simple mathematical calculation determines the position of the extrema at

\begin{equation}
z=\pm \frac{1}{2} \sqrt{\frac{A}{\pi} }= \pm \frac{1}{2} R,
\label{extrema}
\end{equation}
where the voltage takes the value

\begin{equation}
V(\pm \frac{1}{2} \sqrt{\frac{A}{\pi} },A, \chi) = \pm \frac{16 \pi^2 \chi}{ 25 \sqrt{5} A}.
\label{Vextrema}
\end{equation}

The signal, a rise in the voltage, is governed by two parameters the strength $\chi$ and the area of the coil, being proportional to the former and inversely proportional to the latter, thus the ideal radius of the coil $R$ has to be as small as possible compatible with all the approximations.

\begin{figure}[htb]
\begin{center}
\includegraphics[scale= 0.85]{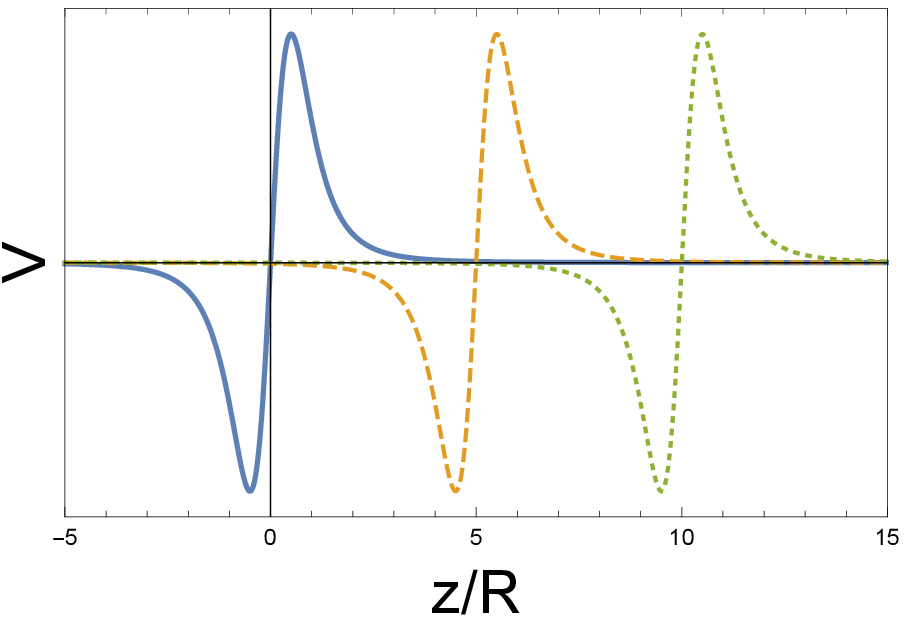} \hspace{1cm} \includegraphics[scale= 0.85]{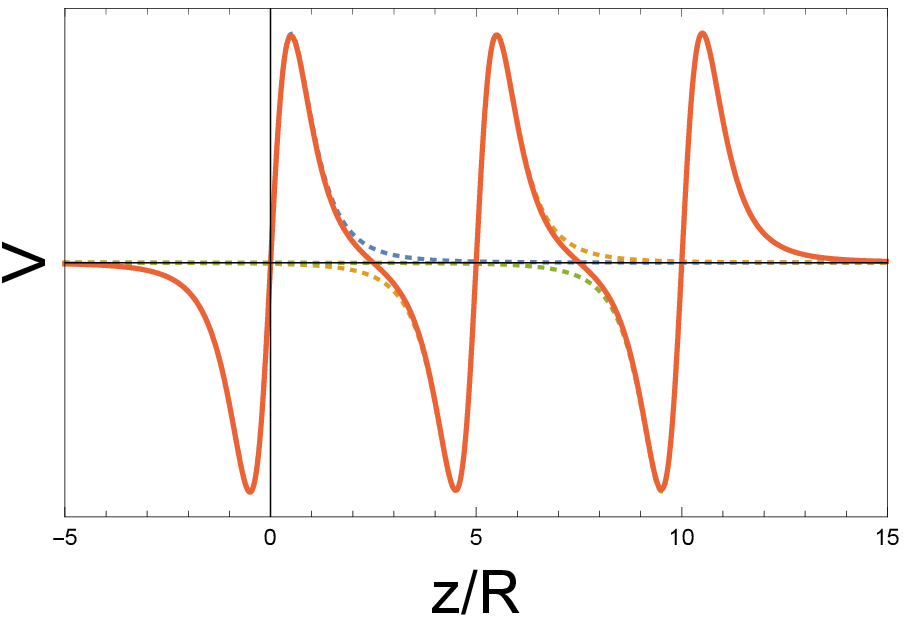} 
\end{center}
\caption{The left figure shows the functional form of the voltage generated by a particle with a magnetic moment when passing three independent coils at uniform speed. The figure on the right shows the combined effect of the three well separated coils forming a solenoid (solid curve) compared with the independent effect of the three coils (dotted curve). The effect is basically the same effect as that of the three coils independently.}
\label{separatedcoils}
\end{figure}

Let us introduce a system with a large number of coils close to each other forming a cylinder. There are two ways of creating this solenoid, either with well separated coils or with closely winded coils. In the first case, as shown in Fig. \ref{separatedcoils},  the effect is simply to add  the effect of each coil almost independently.

\begin{figure}[htb]
\begin{center}
\includegraphics[scale= 0.85]{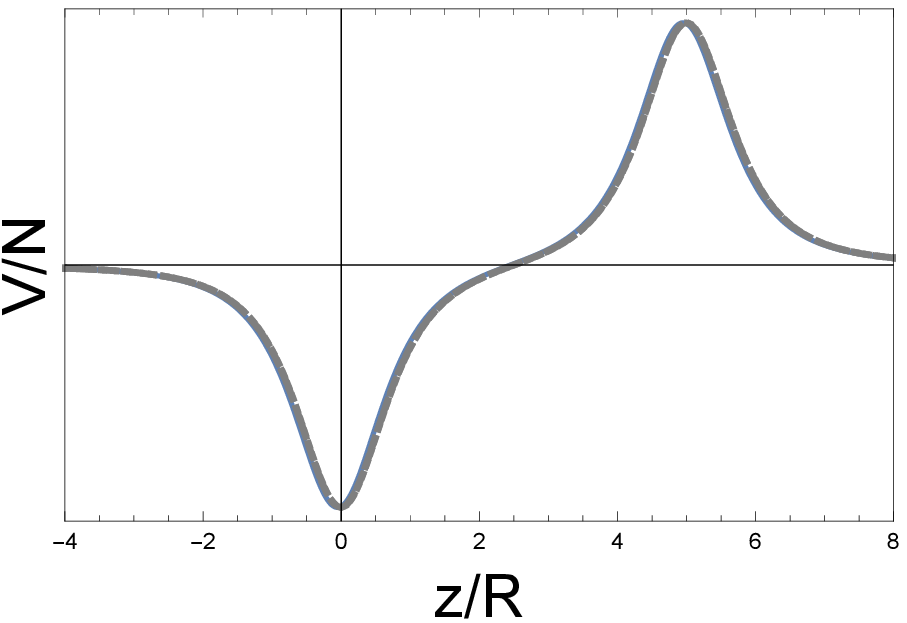} \hspace{1cm} \includegraphics[scale= 0.85]{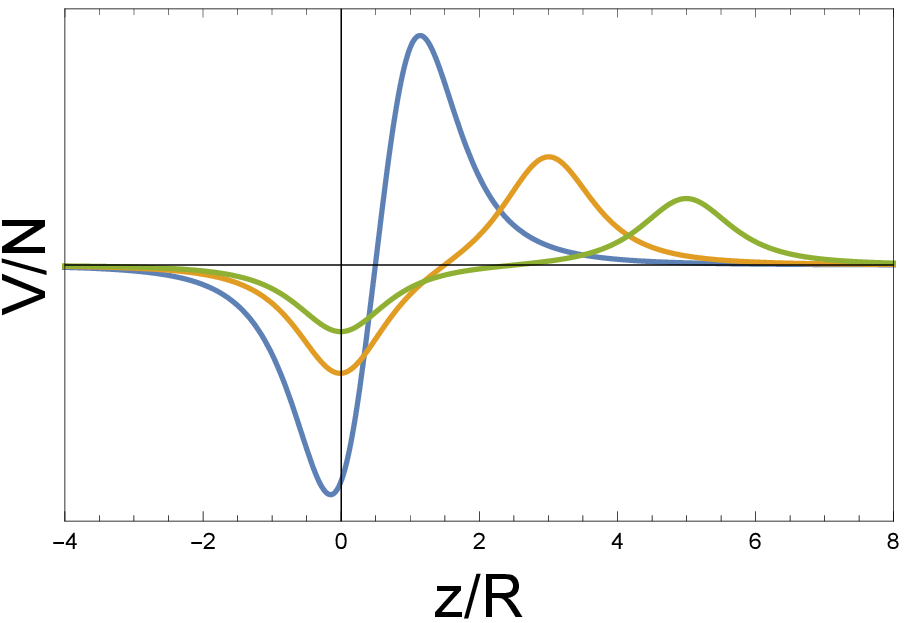} 
\end{center}
\caption{The left figure shows the functional form of the voltage generated by a particle with a magnetic moment when passing a system of conducting coils  of length $5 R$  at uniform speed. The distance between the peaks is the length of the system of coils. The figure is composed by the calculation of the voltage for $50, 500$ and $5000$ loops divided by the number of loops showing that the proportionality between number of loops and voltage persists for relatively large lengths of the systems of coils. The figure on the right shows the same for three different lengths of the coils' system. The shorter the length the greater the voltage.}
\label{faradayN}
\end{figure}

If the coils are very close the effect is more sophisticated the advantage being that we can put many coils in a shorter length to produce an enhancement of the coil effect in shorter distance. This case requires a detailed calculation that can be performed  numerically:

\begin{equation}
V_{N} (z, A, \chi, L) = \sum_{n=1}^{n=N} V(z - (n-1) \Delta z,A, \chi),
\label{exact}
\end{equation}
where $V$ is the potential in Eq.(\ref{emf1}), $N$ is the number of coils, $L$ is the length of the coil cylinder and $\Delta z =\frac{L}{N}$.
We show in  Fig. \ref{faradayN} $V/N$ generated by $50, 500$ and $5000$ coils located between $0\le\frac{z}{R}\le5$. This result is only valid for a large number of coils, when there is a cancellation between the positive and negative potentials. 

\begin{figure}[htb]
\begin{center}
\includegraphics[scale= 0.9]{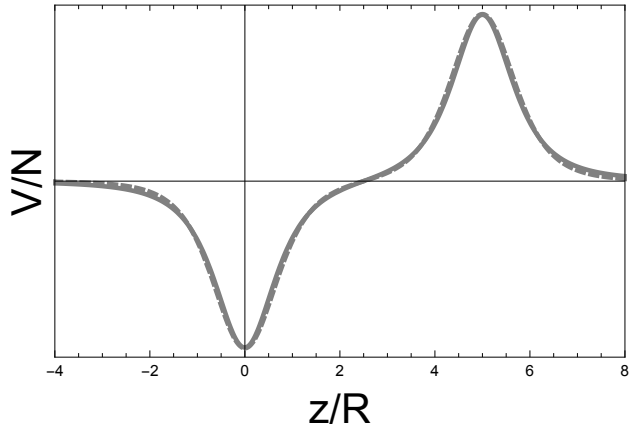} 
\end{center}
\caption{Functional form of the exact numerical calculation Eq.(\ref{exact})(solid) and the approximate one Eq.(\ref{solenoid})(dashed) for $N=1000$, $L = 5 R$  $A=0.1 R^2 $ and $\chi = 0.5$ V$\cdot R$.}  
\label{approximation}
\end{figure}

Doing some numerical analysis we find the relation between the maxima of the single coil potential and the $N$ coil potential for $N$ large

\begin{equation}
\frac{V_{N} (max)}{N} \sim 0.1312  \sqrt{A} \,V_1(max) \sim 0.3706 \frac{\chi}{\sqrt{A}}.
\end{equation}
Moreover, we are able to find an approximate analytical solution valid for large $N$

\begin{align}
\frac{V_{N \, approx}^{in}(z)}{N} &\sim \frac{ -b}{(z^2 + \frac{a}{\pi})^{\sfrac{5}{2}}}, \nonumber\\
\frac{V_{N \, approx}^{out} (z)}{N}& \sim \frac{ b}{((z - L)^2 + \frac{a}{\pi})^{\sfrac{5}{2}}}, \nonumber \\
&
\label{solenoid}
\end{align}
where $a = 2 A$ and $ b= 0.1198\, \chi \,A^2$. The $in$ potential describes the approximation to the  incoming minimum and the $out$ potential the approximation to the outcoming maximum. In Fig.\ref{approximation} we show the result of the exact calculation for $N=1000$ and compare it to  the approximation for the mentioned values of the parameters. 

It is interesting to note from the above expression that the peak of the potential depends directly on the magnitude of the magnetic dipole moment and is inversely proportional to the square root of the area of the coil. This means that one does not need large coil sizes to reveal the magnetic dipole moment. Certainly there are limitations to the sizes due to the approximations used, i.e for circular coils $A= \pi R^2$  and $R$ has to be much bigger than the section of the coil. The result of the calculation shows that the voltage scales nicely with the number of coils and that the smaller the coil system the higher the potential for the same number of coils. One has thus to find a compromise between length and number of coils. The width of the coil has not been taken into account in this calculation, but its effect is irrelevant at this point as long as its size $s$ is smaller than $R$.

\section{Determination of observables characterizing  particles with magnetic moment and no electromagnetic charges}
\label{Sdetector}

In the first subsection we will calculate the current created in systems of coils by the moving dipole as a means of characterizing particles with magnetic moment and no charge. In the second subsection we will study the same effects on particles with magnetic moments of finite lifetime. In the third subsection we study the effect on the particle motion of this induced current which turns out to be small.

\subsection{The induced current and the deposited energy in coil systems}

A dipole traversing a single coil  creates a magnetic field due to the ${\overrightarrow{\cal M}}\cdot \hat{k}$ component of its magnetic dipole moment whose flux  in the coil generates, via the Faraday-Lenz law, an EMF Eq.(\ref{emf}),

\begin{equation}
V_1 (z) = 6 {\mathcal M}_z \, \gamma \,v \, A\,z  \frac{1}{(z^2 + \frac{A}{\pi})^{\sfrac{5}{2}}}.  \\
\label{emf2}
\end{equation}

Only the $z$ component of the magnetic moment  $ {\overrightarrow{\cal M}} \cdot \hat{k}  = {\cal M}_z = {\cal M}\cos{\theta}$, where $\cos{\theta} =\frac{{\overrightarrow{\cal M}} \cdot \hat{k}}{{\cal M}}$ is active. Since we expect the dipoles initially in random directions  in the forward direction, the average ${\cal M}_z$ will be {\cal M} multiplied by the factor, 

\begin{equation}
\frac{\int_0^{2\pi} d\varphi \int_{-\frac{\pi}{2}}^{\frac{+\pi}{2}} d \theta \cos{\theta}}{\int_0^{2\pi} d\varphi \int_{-\frac{\pi}{2}}^{\frac{+\pi}{2}} d \theta } = \frac{2}{\pi}.
\end{equation}
and therefore Eq.(\ref{emf2}) becomes

\begin{equation}
V _1(z) = \frac{12}{\pi} {\mathcal M}\, \gamma \,v \, A\,z  \frac{1}{(z^2 + \frac{A}{\pi})^{\sfrac{5}{2}}} .\\
\label{coil1}
\end{equation}

For the $N$ coil solenoid, recalling Eqs.(\ref{solenoid}), the EMF becomes

\begin{align}
\hspace{2.2cm}V_{N }^{in\;} (z) &\sim -0.4576\, {\mathcal M} \, \gamma \,v \, N\, A^2  \frac{1}{(z^2 + \frac{2 A}{\pi})^{\sfrac{5}{2}}}, \nonumber \\
V_{N }^{out } (z)& \sim +0.4576\, {\mathcal M} \, \gamma \,v \,N\, A^2 \frac{ 1}{((z - L)^2 + \frac{2 A}{\pi})^{\sfrac{5}{2}}},  \nonumber\\
&
\label{solenoid1}
\end{align}
where $N$ is the number of coils in each coil system and $L$ its length.

The total EMF has to take into account the inductance, thus

\begin{equation}
V_T =  - \frac{ d \Phi}{dt} - {\mathfrak L} \frac{d {\mathcal I}}{d t},
\end{equation}
where ${\mathfrak L}$ is the inductance and ${\mathcal I}$ the intensity of the current in the circuit. Ohm's law then becomes the differential equation we have to solve in order to obtain the intensity in the circuit

\begin{equation}
\frac{ d \Phi}{dt} + {\mathfrak L} \frac{d {\mathcal I}}{d t} + {\mathcal I } {\mathcal R} = 0.
\label{Ohm}
\end{equation}
where  ${\mathcal R} = \frac{2 \pi R}{S} \varrho N$, where $S$ is the section of the conductor and $\varrho$ the resistivity which changes from conventional conductors, $\varrho= 10^{-8}$ Ohm$\cdot$m,  to superconductors $\varrho= 10^{-26}$ Ohm$\cdot$m. 

If $\mathcal R $ corresponds to a conductor the inductance term is small and only the induced EMF enters the equation, thus the intensity becomes ${\mathcal I }(z)  \approx - \frac{1}{\mathcal R} \frac{d \Phi}{dt} = \frac{V(z)}{\mathcal R} $ and therefore from Eqs.(\ref{coil}) and (\ref{solenoid1}) we get for one coil

\begin{equation}
{\mathcal I}_1(z) = \frac{12}{\pi} \frac{{\mathcal M}\, \gamma \,v \, A\,z }{ {\mathcal R}}\frac{1}{(z^2 + \frac{A}{\pi})^{\sfrac{5}{2}}} = \frac{12}{\pi} \frac{{\mathcal M}\, \gamma \,v \, A\,z }{ {\mathcal R}}\frac{1}{(z^2 + R^2)^{\sfrac{5}{2}}}, 
\label{coil2}
\end{equation}
and for the solenoid

\begin{align}
\hspace{1.9cm}{\mathcal I }_N^{in\,}(z)& = - 0.4576 \, \frac{{\cal M}\, \gamma  \,v \, A^2 \, N}{{\mathcal R} } \frac{1}{(z^2 + 2 R^2)^{\sfrac{5}{2}}}, \nonumber\\ 
{\mathcal I }_N^{out}(z)& = + 0.4576 \,\frac{ {\cal M}\, \gamma  \,v \, A^2 \, N}{{\mathcal R}}  \frac{ 1}{((z - L)^2 + 2 R^2)^{\sfrac{5}{2}}}. \nonumber \\
 &
 \label{solenoid2}
 \end{align}
If $ \mathcal R $ corresponds to a superconductor then the term ${\mathcal I}{ \mathcal R}$ is negligible and the intensity is given by ${\mathcal I} (z) = -\frac{\Phi}{{\mathfrak L}}  =  \frac{1} {\gamma\, v\, {\mathfrak L}} \int_{-\infty}^z V (z) dz  $, which becomes for the single coil

\begin{equation}
{\mathcal I}_1 = \frac{4}{\pi} \frac{{\mathcal M} A}{\mathfrak L} \frac{1} {(z^2 +\frac{A}{\pi})^{\sfrac{3}{2}}} = \frac{4}{\pi} \frac{{\mathcal M} A}{\mathfrak L} \frac{1} {(z^2 +R^2)^{\sfrac{3}{2}}},
\label{coil3}
\end{equation}
and for the $N$ coil solenoid

\begin{align}
\hspace{0.8cm}{\mathcal I}_N^{in} (z) & = -\frac{0.4576}{{\mathfrak L}} \, {\mathcal M}\, A^2 \, N \int_{-\infty}^z \frac{1}{(z^2 + \frac{2 A}{\pi})^{\sfrac{5}{2}}} dz   \nonumber \\ &= - \frac{0.3762}{{\mathfrak L}} \, {\mathcal M} \, N \, \frac{z(3 R^2 +z^2)}{(z^2 + 2 R^2)^{\sfrac{3}{2}} }, \nonumber\\
{\mathcal I}_N^{out} (z) & = \frac{0.4576}{{\mathfrak L}} \, {\mathcal M}\, A^2 \, N \int_{-\infty}^z \frac{1}{((z-L)^2+  \frac{2 A}{\pi})^{\sfrac{5}{2}}} dz \nonumber \\&= \frac{0.3762}{{\mathfrak L}} \, {\mathcal M} \, N \frac{(z-L)(3 R^2 +(z-L)^2)}{((z-L)^2 + 2 R^2)^{\sfrac{3}{2}}}. \nonumber \\
&
\label{solenoid3}
\end{align}

Let us study first the case of conducting coils, i.e. ${\mathcal R} \ne 0$ for stable particles. The energy our detector can extract from the flux of dipoles is the integral of the power in time

\begin{equation}
{\cal E}  =  \int_{-\infty}^{+\infty} {\frac{V^2(t)}{{\cal R}} dt }.
\end{equation}
Thus for the single coil the integral is immediate and we get

\begin{equation}
{\mathcal E} = \left(\frac{12}{\pi}\right)^2 \frac{{\mathcal M}^2 \gamma v \, A^2 }{{\mathcal R}} \int_{-\infty}^{+\infty} \frac{z^2}{(z^2 + \frac{ A}{\pi})^5} dz = \frac{45}{8\pi} \frac{ {\mathcal M}^2 \gamma v }{R^3 {\mathcal R}}.
\label{energycoil}
\end{equation} 
For the solenoid the integral expression is

\begin{equation}
{\mathcal E} \sim 0.2094 \frac{ {\cal M}^2 \, \gamma \,v \, A^4\, N^2}{{\cal R}} \int_{-\infty}^{+\infty} \left(\frac{1}{(z^2 + \frac{2 A}{\pi})^{\sfrac{5}{2}} }+ \frac{1}{((z - L)^2 + \frac{2 A}{\pi})^{\sfrac{5}{2}}} \right)^2 dz,
\label{enerysolenoid1}
\end{equation}
and its integration requires some discussion.

The integral of the square terms can be performed analytically while the cross term not. In any case all of the three terms are very convergent and numerically we can get the result quite fast. However, the square terms tend to be much larger than the cross term, specially if $L$ is large compared to the width of the potential bumps, since the cross term only takes into account the overlap between the bumps. This argument has been checked numerically for the scenarios of the present calculation and therefore we neglect safely the cross term and get an analytic result which is 
\begin{eqnarray}
{\mathcal E} &\sim & 0.2094 \frac{ {\mathcal M}^2\,\gamma \,v \, A^4\, N^2}{{\mathcal R}} \frac{35 \pi}{64}\left(\frac{2 A}{\pi}\right)^{-\sfrac{9}{2}} \nonumber \\
&\sim & 2.7451 \frac{ {\mathcal M}^2\,\gamma \,v \, N^2}{{\mathcal R} \sqrt{A} }.
\end{eqnarray}
If we use the following units ${\cal M} $ (fm), $v$ in units of the velocity of light in the vacuum, $S$ (mm$^2$), $R$ (mm) and $\varrho$ (Ohm$\cdot$m) the equation for the energy becomes for the single coil

\begin{equation}
{\mathcal E} = 0.1660 \,10^{-28}\, \frac{ {\mathcal M}^2 \gamma v \, S }{R^4 \varrho } \, \mbox{eV},
\label{energycoil1}
\end{equation}
and for the solenoid

\begin{equation}
\hspace{0.5cm}{\mathcal E} = 0.1455 \, 10^{-29}\, \frac{ {\mathcal M}^2\, \gamma v\, S\, N}{R^2\, \varrho} \, \mbox{ eV}.
\label{normal}
\end{equation}

Let us now study the case of superconducting coils where the inductance term is dominant. In this case

\begin{equation}
{\mathcal E}  = \frac{ {\mathfrak L} }{2} \int_{-\infty}^{+\infty} \frac{ d{ \mathcal I}_N^2(t)}{dt } dt  = \frac{ {\mathfrak L}}{2} \, ( ({\mathcal I}_N^{in})^2(+\infty) + ({\mathcal I}_N^{out})^2(+\infty) )
 =  0.1415\, \frac{{\mathcal M}^2\,N^2}{{\mathfrak L}}
\end{equation}
For a thin long solenoid the ratio of inductance for superconducting and normal solenoids is close to 1 \cite{HIRAKAWA1973287}, thus  $ {\mathfrak L} = \mu_0 \frac{N^2 A}{L}$ where $ {\mathfrak L}$ is in H, if we take $\mu_0 = 4 \pi 10^{-7}$ H/m , $A$ the area of the coil in m$^2$, $L$ the length of the coil in m. Recall that H = Ohm$\cdot$s, then

\begin{equation}
{\mathcal E} = 0.2486 \, 10^{-16} \, \frac{{\mathcal M}^2 L}{R^2} \, \mbox{eV}.
\label{super}
\end{equation}
where we are measuring ${\mathcal M}$ in fm, $R$ in mm and $L$ in m. This choice of three units might seem strange at first, but it is related to the natural scale of the physics and the apparatus: the magnetic moments of the states of monopolium are microscopic ($0-1000$ fm), the most effective coils are of milimeters size ($0.01-1$ mm) , and in order to have very large number of coils we need solenoids lengths of meters.

The behavior of Eq.(\ref{normal}) and Eq.(\ref{super})  are very different. For the former the parameters at our disposal to increase the sensibility of the detector are the number of coils  $N$, the velocity of the dipole $v$, but specially the factor $\gamma$, the radius of the coil $R$ and the resistivity of the material $\varrho$. For the latter the length of the solenoid $L$ and the radius of the coil $R$ but neither the velocity nor the number of coils play a role.

\subsection{Application to particles with finite lifetime}
\label{lifetime}
In the previous subsections  we have considered the effect on coils of permanent magnetic moments. Here we  modify  our formulation to magnetic moments of particles with finite lifetime. By generalizing  Eq.(\ref{energycoil}) to this case we get for the energy stored in one coil during the lifetime of particle 

\begin{eqnarray}
{\mathcal E} & \sim &  \left(\frac{12}{\pi}\right)^2 \frac{{\mathcal M}^2 \gamma\, v \, A^2 }{{\mathcal R}} \int_{-z}^{+z}  \frac{z^2}{(z^2 + \frac{ A}{\pi})^5} dz \nonumber\\
 & \sim & \frac{144\,{\mathcal M}^2 \gamma\, v }{{\mathcal R} R^3} \int_{-a}^{+a} \frac{y^2}{(y^2 + 1)^5} dy \nonumber \\ 
& \sim &0.1353 \,10^{-28}\frac {{\mathcal M}^2 \gamma\, v S }{R^4 \varrho} {\mathcal F}(a)\; \mbox{eV}. 
\label{energycoil2}
\end{eqnarray} 
where $a= \frac{z}{R}$, $z$ is half the distance traveled during the particles lifetime, $v$  
the velocity of the state, $\gamma$ the corresponding  relativistic  factor, ${\mathcal M}$ is the magnetic moment, ${\mathcal R}$ is the resistance of the coil and $R$ the coil radius and $S$ the surface section of the resistor. The integral in Eq.(\ref{energycoil2}) ${\mathcal F} (a)$ can be solved analytically giving

\begin{equation}
{\mathcal F} (a) = \int_{-a}^{+a} \frac{y^2}{(y^2 + 1)^5} dy = \frac{1}{192} \left(\frac{ 15 a^7 + 55 a^5 + 73 a^3 -15 a}{(1+a^2)^4} + 15 \arctan{a} \right).
\label{integral}
\end{equation}
We are assuming that the coil is located in the center of the path traveled by the particle during its lifetime. If this path is large compared with the radius of the coil this is not a bad approximation given the behavior of ${\mathcal F}(a)$. If we put $N$ coils next to each other but sufficiently separated that there is no interaction between them we have to multiply Eq.(\ref{energycoil2})  by $N$.

\begin{figure}[htb]
\begin{center}
\includegraphics[scale= 0.8]{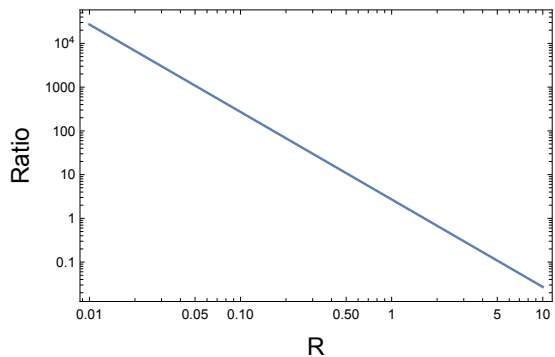}
\end{center} 
\caption{We show the ratio of a coil system with respect to a solenoid. The ratio is almost independent on the number of coils as long as they are the same and $z > R$.}
\label{CoilSolenoidRatio}
\end{figure}

We proceed analogously with the solenoid, and in order to get an analytic formula we make the same approximation as before, namely we disregard the cross term, the result is

\begin{eqnarray}
{\mathcal E} &\sim& 0.2094  \frac{{\mathcal M}^2 \gamma_n\, v_n \, A^4 N^2 }{{\mathcal R}}\left( \int_{-z}^{+z}  \frac{1}{(z^2 + \frac{2 A}{\pi})^5} dz + \int_{-z}^{+z}  \frac{1}{((z-L)^2 + \frac{2 A}{\pi})^5} dz\right) \nonumber \\
& \sim &\;0.1435 \frac{{\mathcal M}^2 \gamma\, v\, S N }{{R^2 \varrho}}\left( \int_{-b}^{+b}  \frac{1}{(y^2+1)^5} dz + \int_{-b}^{+b}  \frac{1}{((y-l)^2 + 1)^5} dz\right) \nonumber \\
& \sim &0.8469 \,10^{-30} \frac{{\mathcal M}^2 \gamma\, v \, S N }{{R^2 \varrho}} ({\mathcal G} (b)  +\frac{1}{2} {\mathcal G} (b - l) + \frac{1}{2}{\mathcal G} (b + l)) \;\mbox {eV}
\label{energycoil3}
\end{eqnarray} 
where $b = \frac{z}{\sqrt{2} R}$, $ l= \frac{L}{\sqrt{2} R}$ and the integrals can be expressed in terms of the function

\begin{equation}
{\mathcal G}(b) = \frac{1}{192} \left(\frac{ 105 b^7 +385 b^5 +511 b^3 + 279 b}{(1+b^4)^2} + 105 \arctan{ b}\right).
\end{equation}
The formulation just described can be applied to any particle with a finite lifetime with magnetic moment and no charge.  

In order to choose the best parameters we show in Fig. \ref{CoilSolenoidRatio} the ratio of the effect of $N$ separated coils with respect to a $N$ coils solenoid for large $N$. The dependence of the coil radius and the coil section is crucial. We choose the radius of the coil section $r_s$ for technical reasons about $r_s \sim 0.01 R$. The figure shows that for large radii $R > 2$ mm the solenoid is more efficient, while for microcoils $R< 2$ mm the coil system is more efficient. We shall use in what follows microcoils $R\sim0.01$ mm and $r_s\sim0.0001$ mm and therefore $N$ coil systems instead of solenoids.

\subsection{Effect of the induced current on the velocity of the particle}

One might wonder if the induced current through the inductance can produce an effect contrary to the one just discussed which reduces the signal significantly. For simplicity we will perform a non relativistic calculation which is very transparent. Let us assume a particle traveling with constant velocity  towards a conducting coil with its magnetic dipole moment perpendicular to the plane of the coil as shown in Fig. \ref{coil}. We would like to find out the motion of this particle due to the effect of the Faraday-Lenz law which creates a magnetic field $\vec{B} = B_z \hat{k}$ where

\begin{equation}
B_z = \frac{2 \pi R^2 {\mathcal I}}{(z^2 +R^2)^{\sfrac{3}{2}}},
\end{equation}
with ${\mathcal I}$ the intensity of the current running through the coil, $z$ is the distance from the particle to the center of the coil and $R$ is the coil radius. Newtons's equation for the problem is given by

\begin{equation}
M \frac{d^2z}{dt^2}= - {\mathcal M} \frac{d B_z}{dz}.
\end{equation}

\begin{figure}[htb]
\begin{center}
\includegraphics[scale= 0.58]{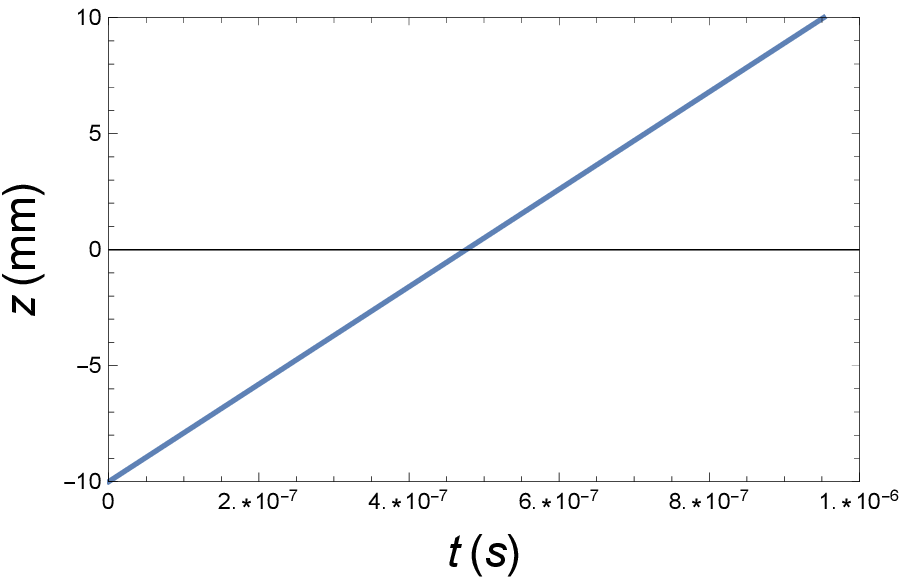} \hspace{0.1cm}\includegraphics[scale= 0.67]{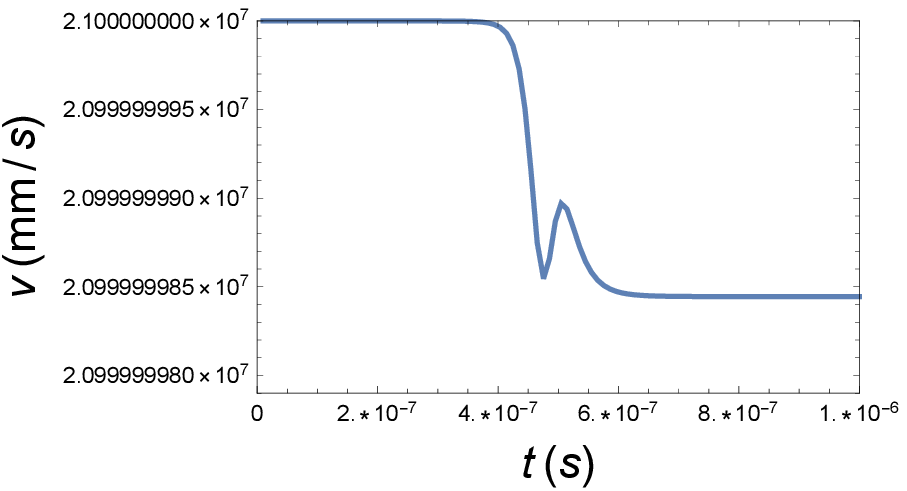}\hspace{0.1cm}\includegraphics[scale= 0.6]{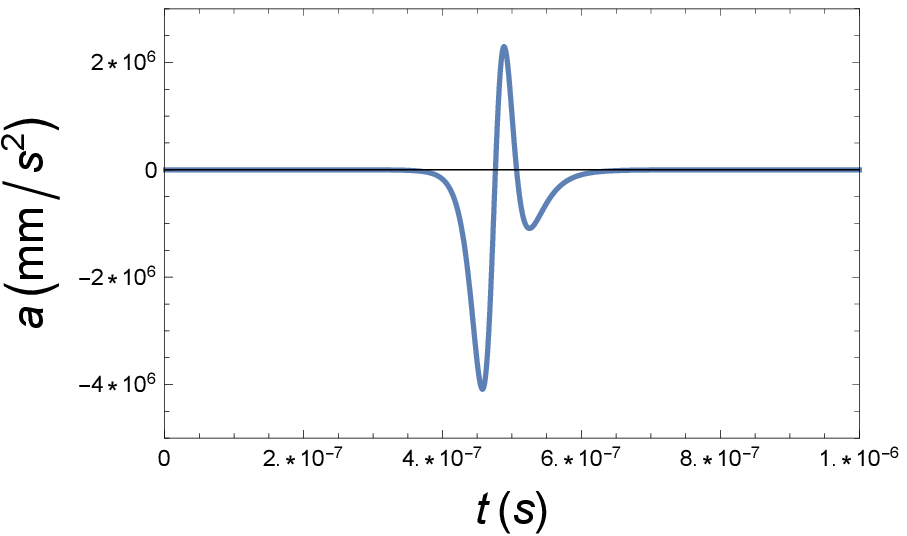}  \\[0.2cm]
\includegraphics[scale= 0.6]{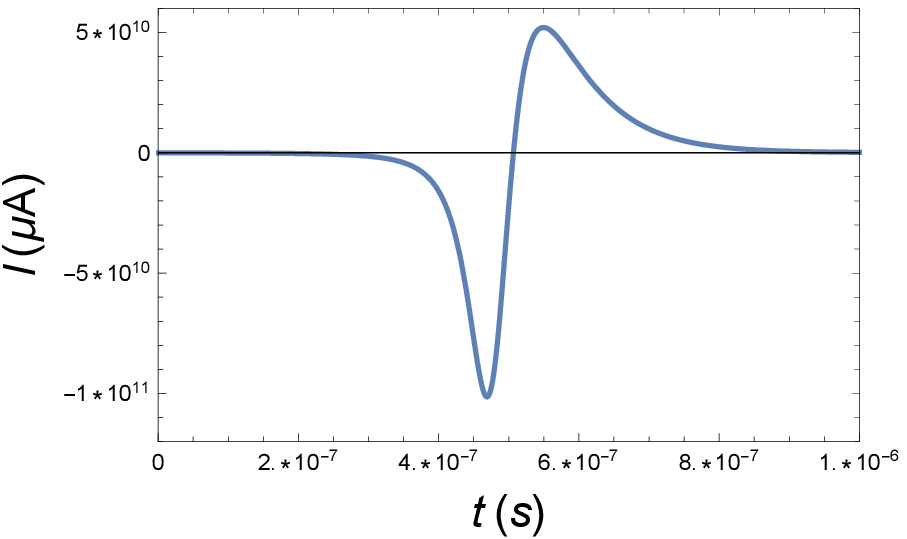} \hspace{0.2cm}\includegraphics[scale= 0.6]{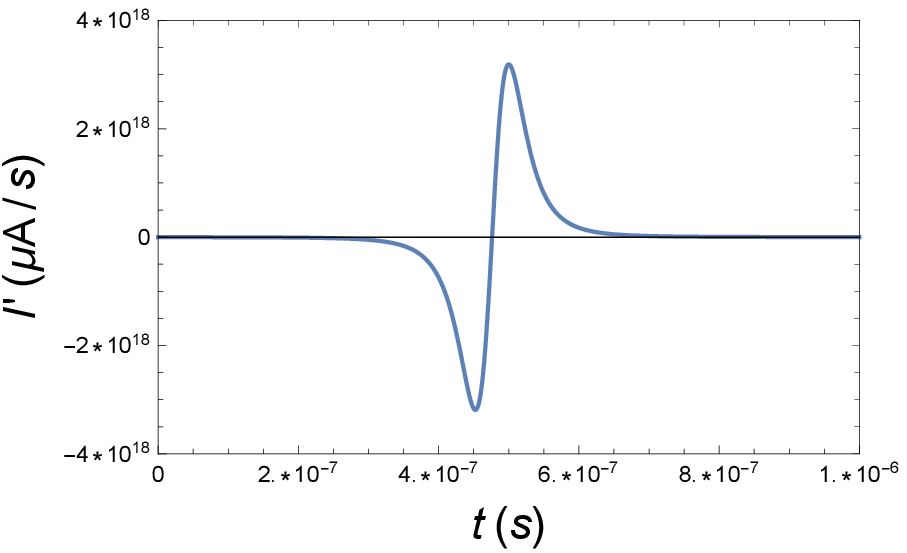}
\end{center} 
\caption{A high excited monopolium (${\mathcal M} \sim 300 $ fm) traveling  with constant velocity ($v=7\;10^{-5}$) from $z_i= -10$mm towards a conducting coil located at $z=0$ with its magnetic dipole moment perpendicular to the plane of the coil. The upper figures from left to right show variation of the position, velocity and acceleration with time. The lower figure show the variation of the induced current  and its derivative with time.}
\label{FuerzaEspiraEx}
\end{figure}

In order to get the intensity we use Ohm's law Eq.(\ref{Ohm}) and and the magnetic flux $\Phi$ given by Eq.(\ref{flux1}).
Since we are dealing with a very good conductor, 
${\mathcal R} = \frac{2 \pi R}{\pi r_s^2} \varrho $, where $r_s$ is the radius of the conductor and $\varrho$ the resistivity  $\varrho= 10^{-8}$ Ohm$\cdot$m and has a small  inductance ${\mathcal L} =0.1$ nH. Thus, the set of equations that determine the motion of the particle are

\begin{eqnarray}
M \frac{d^2 z}{d t^2} & = & - 6 \pi {\mathcal M} R^2 {\mathcal I} \frac{z}{(z^2 +R^2)^{\sfrac{3}{2}} },\\
 {\mathcal L} \frac{d {\mathcal I}}{d t} + {\mathcal I } {\mathcal R} &= & 6 \pi {\mathcal M} R^2 {\mathcal I} \frac{z}{(z^2 +R^2)^{\sfrac{3}{2}} }\frac{d z}{dt}.
\end{eqnarray}
This is a system of differential equations for $z(t)$ and ${\mathcal I}$ (t) which has to be solved simultaneously. The initial conditions will be $z_i ,v_i$ and ${\mathcal I}_i$. We shall take $z_i$ sufficiently far from the coil so that the induced current is very small initially. The parameters that influence the solution  are ${\mathcal L}, {\mathcal R}, {\mathcal M}$ and $v_i$.  The outcome of our calculation is shown in Fig. \ref{FuerzaEspiraEx}.

As can be seen in Fig. \ref{FuerzaEspiraEx}, the particle approaches the coil at large distances without changing its velocity. Close to the coil, for large ${\mathcal M}$, the velocity drops due to an acceleration opposite to the direction of motion associated to the creation of an induced current. Due to the variation of the current the inductance term becomes active and generates an acceleration opposite to the one generated before, which increases the velocity slightly and therefore decreases the current. Finally the Ohm term dominates over the inductance term and the velocity is decreased again and the current increases, until the particle leaves the region of influence of the coil with a smaller velocity and stops to generate current. The figure corresponds to the solution of the equations for $z_i=-10 $mm, $v_i= 7.\,10^{-5}$, $t= 10^{-6}$s, and a very high lying Rydberg state with  ${\mathcal M} \sim 300$ fm. The effect of the inductance term on the velocity is very small,  and the full change of the velocity by the coil is also small so that the position has a negligible kink at the origin. In this case the velocity changes from $2.1 \, 10^7$ mm/s to $2.099999985\,10^7 $ mm/s and if we follow the initial trajectory beyond the coil $z_-(0.6\,10^{-6}) =2.6$ cm, while the position the particle follows after the coil is $z_+(0.6\,10^{-6}) =2.59999998$ cm. 

The conclusion of this analysis is twofold: on the one hand the induced current does not influence significantly the calculation and from now on we shall not take the induction effect into consideration; on the other hand the effect of the coil on the velocity is small. Despite this, given the large mass of the monopolium states, the created current is not negligible.

\section{Analysis of monopolium properties using coils and solenoids}
\label{monopolium}

We proceed to apply the coil effect to  monopolium a monopole-antimonopole bound state and its excitations.  We proceed to discuss  two experimental scenarios of monopolium:  one related to a relatively light monopolium ($M \sim 2000$ GeV) which can be created by particle collisions~\cite{Epele:2012jn,Mavromatos:2020gwk} and one related to a heavy monopolium ($ M\sim M_{GUT}$ and $M \sim M_{Planck}$) that might be created in cosmological scenarios~\cite{Preskill:1984gd,Vento:2020vsq}.

 \subsection{Low mass monopolium}
\label{lowmassmonopolium}

Let us review some properties of monopolium which are useful for the calculation of low mass monopolium. There are several models in the literature to describe the  monopole-antimonopole potential in monopolium  \cite{Goebel1970,PhysRev.160.1257,Barrie:2016wxf,Baines:2018ltl,Barrie:2021rqa}. For the purpose of our present investigation the analytic approximation to the potential of Schiff and Goebel \cite{Goebel1970,PhysRev.160.1257} 

\begin{equation}
V(r) = -g^2 \frac{1-exp(-2 r/r_0)}{r},
\label{Goebel}
\end{equation}
used in refs.~\cite{Epele:2007ic,Epele:2008un} will be sufficient. The approximation consists in substituting the true wave functions by Coulomb wave functions of high $n$. For each $r_0$ in Eq.(\ref{Goebel}) a different value of $n$ will be best suited. We use the equation

\begin{equation}
\rho_n= 48 \alpha^2  n^2,
\label{size}
\end{equation}
to parametrize all expectation values in terms of $\rho_n$, where $\rho_n=r_M/r_{classical}$, $r_M$, being the radius of the monopolium state,  $r_{classical}$ is the classical magnetic radius of the monopole  $g^2/m$,  $\alpha$ is the electromagnetic fine structure constant. This relation results from the calculation of the expectation value of $r$ in the $(n,0)$  Coulomb state. 

We determine an approximate wave function for the ground state of the Schiff-Goebel potential in terms of a Coulomb wave function with a $n_{min}$. In terms of $\rho_{n_{min}}$ the mass of ground state monopolium becomes~\cite{Epele:2007ic,Epele:2008un}

\begin{equation}
M=  m\left(2-\frac{3}{4 \rho_{n_{min}}}\right).
\label{binding}
\end{equation}
Thus given a monopole mass $m$ and a monopolium ground state mass $M$ we determine $\rho_{n_{min}}$ which determines $n_{min}$. The binding energy of the ground state is thus $E_{n_{min}} = \frac{3 m}{4 \rho_{n_{min}}}$. The excited states correspond to $n> n_{min}$. Let us label the excited states only by their principal quantum number $n$ since we are assuming in our simple Coulomb potential model that all the $l$ corresponding to an $n$ are degenerate, thus  their mass would be

\begin{equation}
M_n=  m\left(2-\frac{3}{4 \rho_n}\right),
\label{binding1}
\end{equation}
and their binding energy $E_n = \frac{3 m}{4 \rho_n}$. The high lying deformed  ($l\ne 0$) states will have a larger magnetic moment and a larger size. Recall that in a Coulomb like potential

\begin{equation}
\frac{\langle n  l | r | n  l \rangle}{r_{\mbox{{\tiny Bohr}}}} \sim (\frac{3}{2} n^2 -\frac{1}{2}{  l}({ l}+1))  \sim  n^2,
\end{equation} 
an approximation valid for large $l$. Thus the high lying states will increase their size approximately as $n^2$. We estimate their magnetic moment as,
 
\begin{equation}
{\mathcal M_n} \sim  g \, \langle n l | r | n l\rangle = g\, \rho_n \,r_{classical} =  \frac{3}{32\,\alpha^{\sfrac{3}{2}}\,E_n},
\label{magneticmoment}
\end{equation}
where we have used the Dirac quantization condition (DQC) $g \,e =\frac{1}{2}$ and $r_{classical} = g^2/m$.  Looking at Eq.(\ref{magneticmoment}) it is clear that Rydberg states will have the largest magnetic moments. 

The two photon decay width of the monopolium excited states is determined by~\cite{Epele:2008un,Epele:2012jn} 

\begin{equation}
\Gamma^{2ph}_n(E) = \frac{2\,\beta_n^4\, m^3}{\alpha^2 M_n^2} \left(\frac{E_n}{m}\right)^{\sfrac{3}{2}}.
\label{twophoton}
\end{equation}
At leading order the dominant decay rates are determined by the two photon width, Eq.(\ref{twophoton}), which depends on the wave function at the origin, and is only non zero for S ($l=0$) states. The $l\ne 0$ states arise by de-excitation to lower $n$ levels. The decay rate for the S states goes like $1/n^3$  \cite{Dirac:1930bga,Ore:1949te}.

Our interest here lies in the high lying deformed Rydberg states which are in high $l$ states, with vanishing wave function at the origin, that is to say that the monopole and antimonopole are far away from each other and therefore do not annihilate as it is the case for positronium. They will de-excite by spontaneous emission of photons to the lower levels where they will annihilate either as S states or to non leading order  as $l \ne 0$ state decays. In order to get an estimate for spontaneous emission as a function of $n$ we use the naive Bohr model ~\cite{Dicus:1983ry},

\begin{equation}
\Gamma^{se}_n(E) = 4096 \,\beta_n^4 \, \alpha \,m \left(\frac{E_n}{m} \right)^3.
\label{spontaneousemission}
\end{equation}
Note that this dependence goes like $1/n^6$, therefore lifetimes of these Rydberg states increase notably with $n$.

Two coupling schemes have been used in the literature: the Dirac coupling scheme~\cite{Dirac:1931qs,Dirac:1948um} is characterized by point like monopoles which couple with magnetic charge $g$, while the $\beta$ or Schwinger scheme~\cite{Schwinger:1976fr} is characterized by a monopole coupling of magnetic charge $\beta g$, where $\beta$ is the velocity of the monopole. In Eqs. (\ref{twophoton}) and (\ref{spontaneousemission}) $\beta_n=1$ for Dirac coupling and $\beta_n=v_n$ for Schwinger coupling,  where $v_n$ is the velocity of the monopolium state, $\alpha$ is the fine structure constant, which appears after using the DQC and  $m$ stands for the monopole mass.

Let us next specify the properties of monopolium excited states when passing through coils in each of these schemes.

\subsubsection{The Dirac coupling scheme}
\label{Dirac}

\begin{figure}[htb]
\begin{center}
\includegraphics[scale= 0.95]{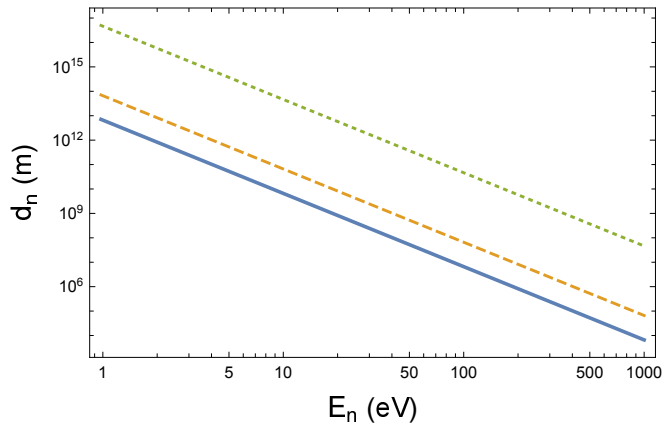} 
\end{center}
\caption{Distance in m traveled by monopolium with binding energy $E_n$ in eV before decaying for three different velocities $0.001$ (solid),$ 0.01$ (dashed) and $ 0.99$ (dotted).}
\label{distanceD}
\end{figure}

The states with the largest magnetic moment, i.e. the most deformed states will be Rydberg states with very small binding energy, $E_n << 2m$. The distance traveled by monopolium with binding energy $E_n= \frac{m}{64\alpha^2 \,n^2}$ for high $n$ before disintegrating is

\begin{equation}
d_n(E_n^{tot}) = \frac{\gamma_n\,v_n} {\Gamma^{se}_n} \sim\frac{\gamma_n\,v_n} {4096 \, \alpha \,m \left(\frac{E_n}{m} \right)^3}
\sim 64 \frac{v_n}{\sqrt{1-v_n^2}}\,\frac{\alpha^5}{m} \,n^6,
\label{distancetraveled}
\end{equation}
since spontaneous emission dominates to annihilation for high $n$. $E_{n}^{tot}= \gamma_n\,M_n$, where $M_n$ is the mass of the state, is the total energy of the monopolium state. We plot the distance $d_n (E_n^{tot})$ in Fig. \ref{distanceD} for monopoles of mass $m \sim 1000$ GeV as a function of $E_n$ for Rydberg states. The curve diverges for $E_n\rightarrow 0$, i.e. $n \rightarrow \infty$ indicating that for small binding energies the distance traveled by monopolium before decaying is very large.  

\begin{figure}[htb]
\begin{center}
\includegraphics[scale= 0.95]{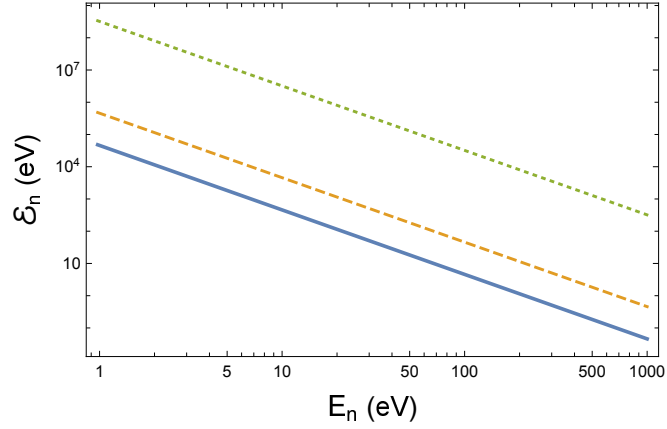} 
\end{center}
\caption{Energy deposited in coil system with $10^8$ coils by the passing of a monopolium state $n$ formed by monopoles mass $m=1000$ GeV with three velocities (0.001 (solid), 0.01 (dashed), 0.99 (dotted)) as a function of binding energy $E_n$ (eV) for the Rydberg states.}
\label{energycoilD}
\end{figure}

Let us now look at our signal which is the deposited energy. In Fig. \ref{energycoilD} we show ${\mathcal E}_n$ for $10^8$ coils as a function of the binding energy  $E_n$ in eV for $\mathcal R = \frac{2\pi R {\varrho}}{\pi r_s^2}$ with the radius of the coils $R=0.01$ mm, the radius of the section of the coil $r_s=0.0001$ mm, the resistivity ${\varrho}=10^{-8}$ Ohm$\cdot$m and for three velocities $0.001, 0.01$ and $0.99$. The coil system is defined by coils separated $2r_s$ from the next coil. $10^8$ coils define a coil system of $40$ m. The  figure shows that the deposited energies for small bindings are measurable.

\subsubsection{The Schwinger coupling scheme}
\label{Schwinger}

\begin{figure}[htb]
\begin{center}
\includegraphics[scale= 0.85]{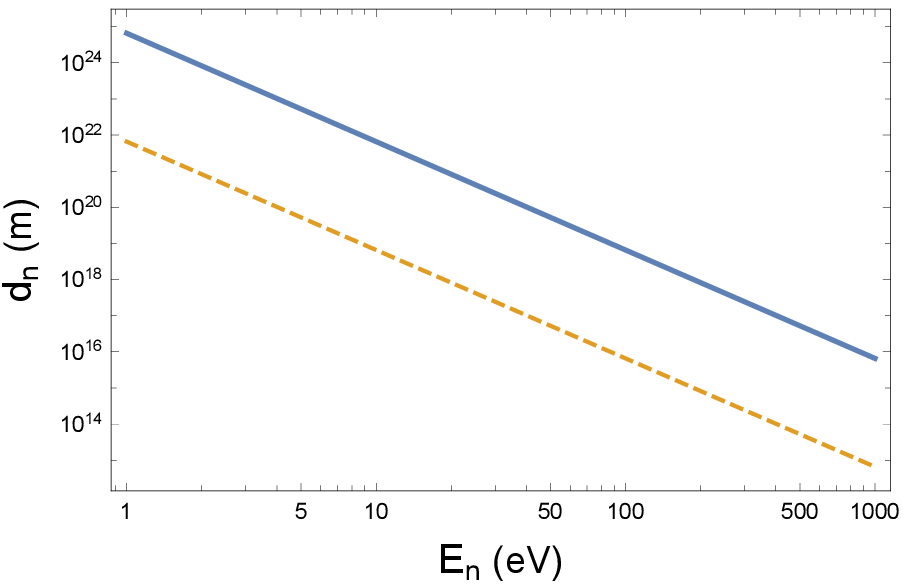} \hspace{0.5cm} \includegraphics[scale= 0.85]{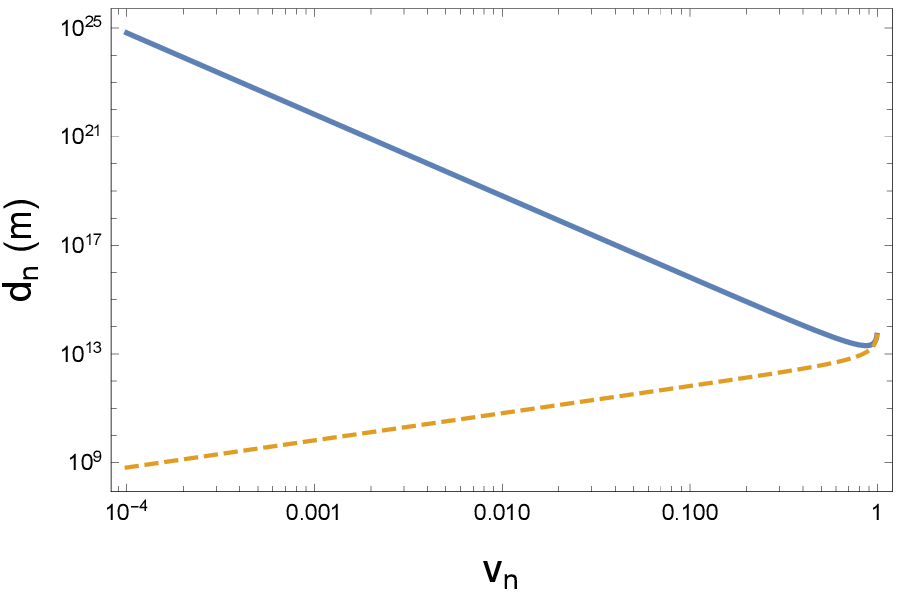} 
\end{center}
\caption{The right figure shows the distance traveled by an excited state of monopolium as a function of binding energy for Schwinger coupling close to threshold $v_n \sim 0.001$ (solid) and $v_n= 0.01$ (dashed). The left figure shows the distance traveled by an excited state of monopolium of binding energy $E_n= 10$ eV as a function of the state's velocity both for Schwinger coupling(solid) and for Dirac coupling (dashed) which demonstrates clearly the threshold effect.}
\label{distanceS}
\end{figure}

The velocity in Eq.(\ref{spontaneousemission}) in Schwinger coupling makes the lifetime very long at threshold and therefore we expect some interesting physics around the threshold region. Away from threshold both Dirac and Schwinger coupling schemes coincide. Let us calculate the value  of $d_n$ at threshold

\begin{equation}
\lim_{E^{tot}_n \rightarrow M_n} d_n (E^{tot}_n) \sim \frac{m^2}{4096\, \alpha \,E_n^3} \left(\frac{M_n}{2 (E^{tot}_n - M_n)}\right)^{\sfrac{3}{2}} \sim \frac{m^2}{4096\, \alpha \,E_n^3 \,v_n^3} 
\end{equation}
The distance traveled by monopolia diverges at threshold as shown in Fig. \ref{distanceS}. The same figure also compares the results for Schwinger and Dirac schemes, noting that close to threshold the lifetimes are very different while as the velocity increases they become similar.

The threshold behavior for the deposited enegy can be calculated analytically from Eqs.(\ref{energycoil2}) and (\ref{integral}) using that at threshold $a_n \sim \frac{d_n}{2\,R} \rightarrow  \infty$, therefore ${\mathcal F}(a_n) \rightarrow \frac{5 \pi}{128}$, we get

\begin{equation}
\lim_{E\rightarrow M_n}  {{\mathcal E}_n}\sim 0.1353 \,10^{-28} \frac{{\mathcal M }_n^2 \gamma_n v_n S N}{R^4 \varrho} {\mathcal F}(a_n) \sim 0.2348\, 10^{-29} \frac{{\mathcal M }_n^2 S N}{R^4 \varrho} \left( \frac{E^{tot}_n - M_n}{E^{tot}_n}\right)^{\sfrac{1}{2}} \sim 0.1660\;10^{-29} \frac{{\mathcal M }_n^2 S N}{R^4 \varrho} v_n \;\;  \mbox{eV}.
\label{thresholdlimit}
\end{equation}

\begin{figure}[htb]
\begin{center}
\includegraphics[scale= 0.85]{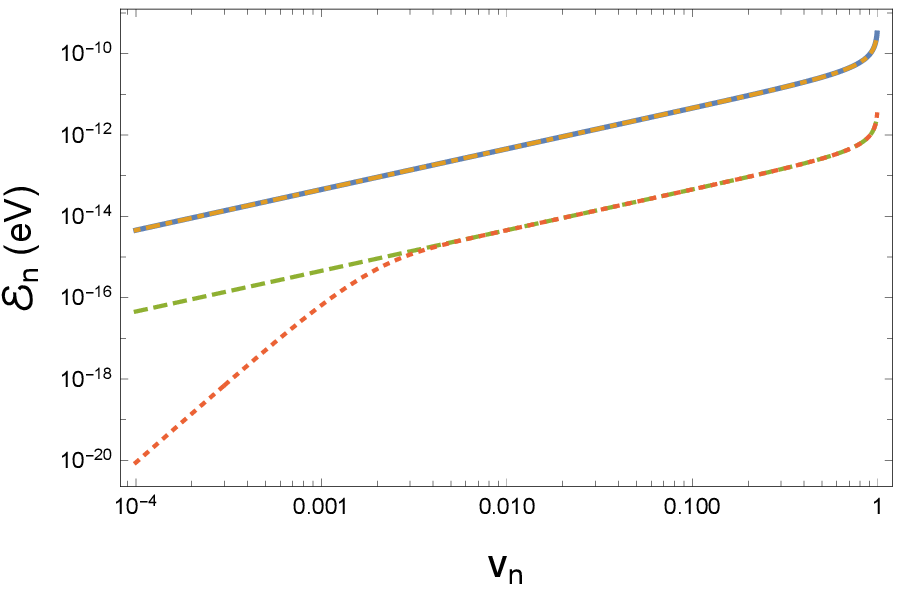} \hspace{0.5cm} \includegraphics[scale= 0.85]{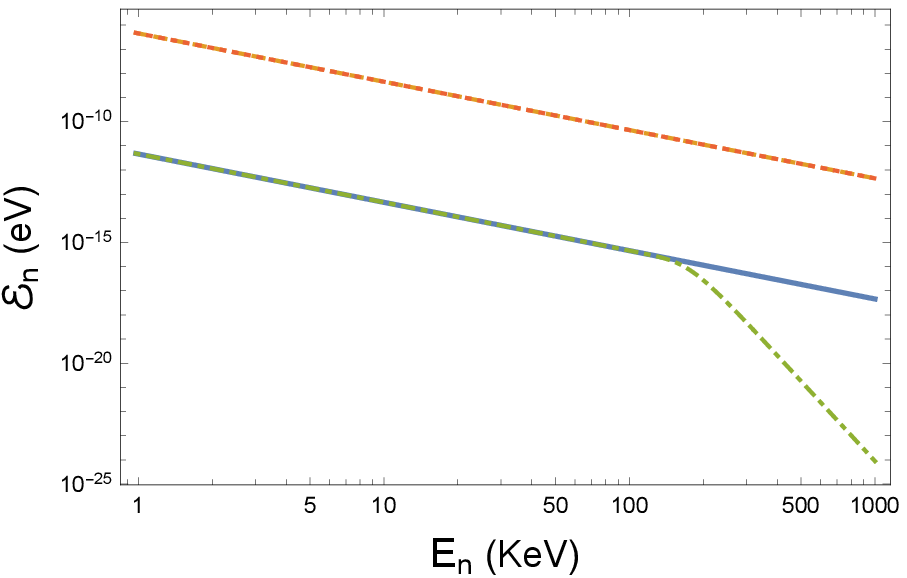} 
\end{center}
\caption{Left: We show the variation of the deposited energy with velocity in one coil for fixed binding energy $En=100$ KeV in the Schwinger scheme (solid) and in the Dirac scheme (dotdashed) and for  $E_n =1000$ KeV in the Schwinger scheme (dashed) and in the Dirac scheme (dotted).  Right:  We show the variation of the deposited energy in one coil with binding energy $E_n$ for fixed velocity $v_n=0.00001$ close to threshold  in the Schwinger scheme (solid) and in the Dirac scheme (dotdashed) and for  $v_n =0.7$ away from threshold in the Schwinger scheme (dashed) and in the Dirac scheme (dotted).}.
\label{DS}
\end{figure}

This linear behavior with the velocity appears clearly in the exact numerical calculations of Fig. \ref{DS}, where we show the behavior of the deposited energy for fixed binding energy and varying velocity (left) and for fixed velocity and varying binding energy (right) in both Schwinger and Dirac schemes. We note that in the Schwinger scheme the linear velocity dependence ceases at large velocities, while in the Dirac scheme the behavior at threshold is not linear. However, for small binding energies the behavior of the two schemes is dominated by the binding energy  whose behavior is the same. This leads to the superposition of the  solid curve  to the dotdashed  curve for small binding in the figure on the left and for small binding and small velocity in the figure on the right. For high binding energies and small velocities  the two schemes tend to deviates considerably as can be seen in the figure on the right. Note that in the Schwinger scheme also the deposited energy grows as the velocity increases as happens in the Dirac scheme.

\begin{figure}[htb]
\begin{center}
\includegraphics[scale= 0.85]{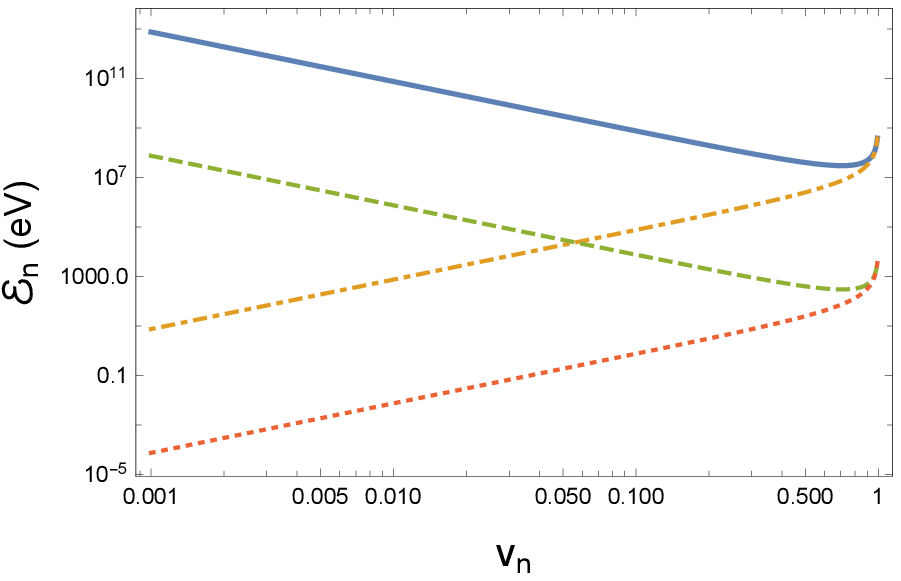} \hspace{0.5cm} \includegraphics[scale= 0.85]{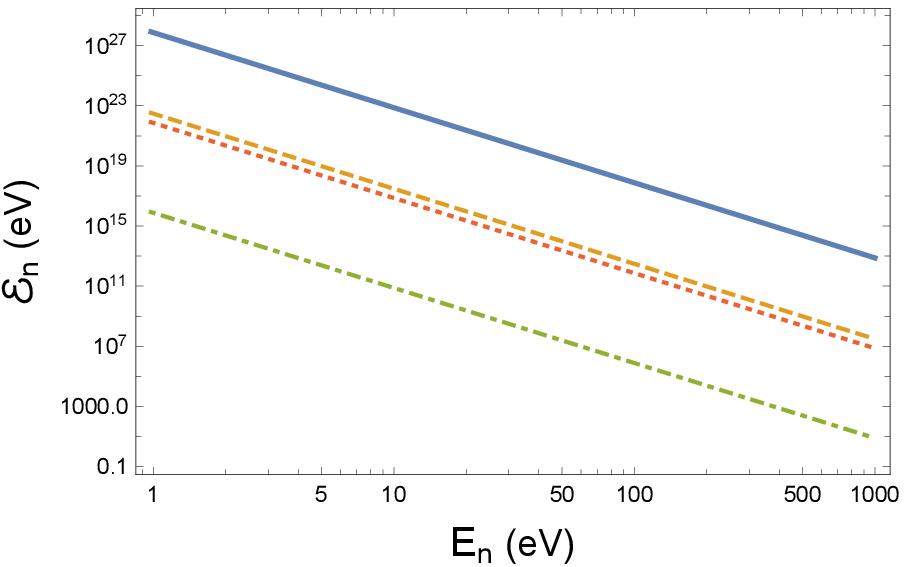} 
\end{center}
\caption{Left: We show the variation of the deposited energy with velocity in the maximal coil system for fixed binding energy $E_n=1$ KeV in the Schwinger scheme (solid) and in the Dirac scheme (dotdashed) and for  $E_n =10$ KeV in the Schwinger scheme (dashed) and in the Dirac scheme (dotted).  Right:  We show the variation of the deposited energy in one coil with binding energy $E_n$ for fixed velocity $v_n=0.001$ close to threshold  in the Schwinger scheme (solid) and in the Dirac scheme (dotdashed) and for  $v_n =0.7$ away from threshold in the Schwinger scheme (dashed) and in the Dirac scheme (dotted).}.
\label{DSN}
\end{figure}

For small binding energies, there is not much difference in the deposited energy for both schemes if both traverse the coil system. The main difference between the Schwinger scheme and the Dirac scheme lies in Fig. \ref{distanceS}, namely that the chance for a Schwinger monopolium to reach a detector is greater because it lives longer and that it can traverse longer coil systems. This behavior is explicit in Fig. \ref{DSN} where we are assuming that monopolium is created at the beginning of the coil system and it traverses a coil system of length $d_n$. In this case we see how the behavior with velocity changes for Schwinger coupling, namely the deposited energy decreases with velocity because the distance traveled by monopolium before decaying becomes smaller, and the detector has less coils. The contrary happens for the Dirac scheme. The behavior with binding energy is the same in both cases, only that the effect for Schwinger coupling is much bigger because the number of coils is larger.

Given the reachable luminosities at LHC, Rydberg states on the scale of eVs, which could lead to measurable deposited energies and lifetimes sufficiently long to escape the beam, cannot be produced and therefore we have to resort to cosmological scenarios \cite{Fanchiotti:2021xyj}.

\section{Cosmic monopolium}
\label{cosmic}

\subsection{Cosmic Light Monopolium}
\label{light monopolium}

\begin{figure}[htb]
\begin{center}
\includegraphics[scale= 0.8]{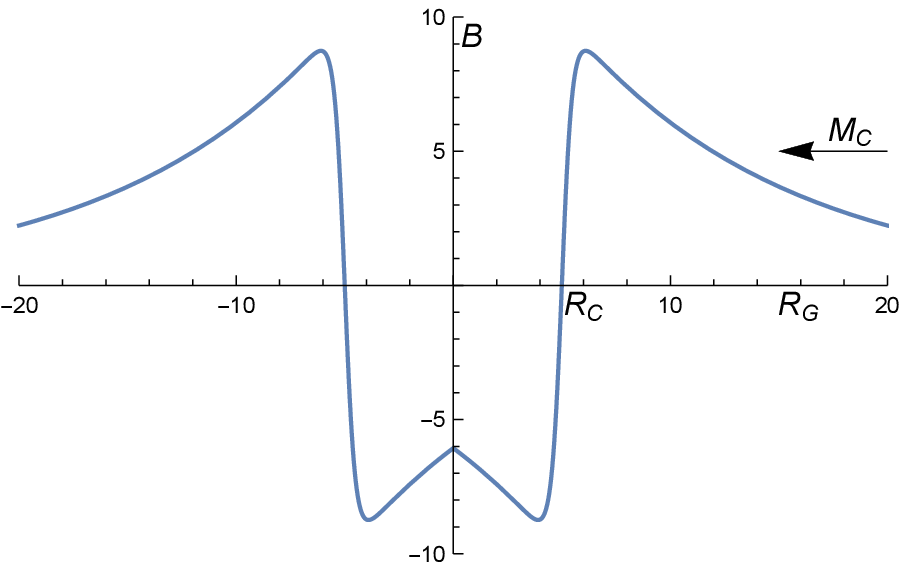} \hspace{1cm} \includegraphics[scale= 0.8]{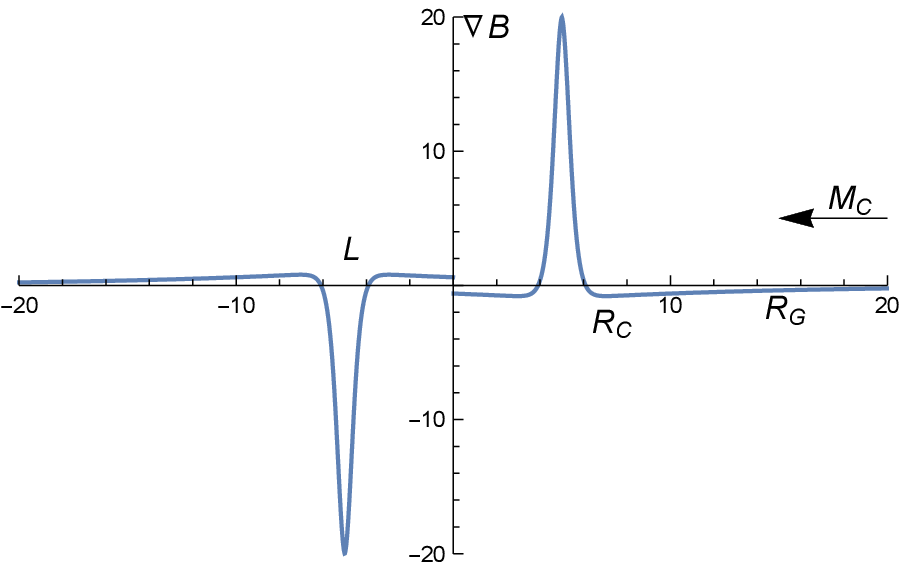} 
\end{center}
\caption{Figure left: our one dimensional model for the magnetic field of a galaxy. Figure right:  magnetic field gradient of the model. $M_C$ represents a monopolium cloud traveling towards the center of the galaxy.}
 \label{galaxy}
\end{figure}
In the case of the monopolia just studied, i.e. light monopolia ($M_n \sim 2000$ GeV), we have to generalize the calculation of their velocity in a galaxy and find a reasonable flux bound. Monopolia will interact with the cosmological magnetic field through their magnetic moment, thus the force acting on them will be given by

\begin{equation}
\vec{F} = - \vec{{\mathcal M}} _n\cdot \vec{\nabla} \vec{B},
\end{equation}
where ${\mathcal M}_n\sim g r_n $ is the magnetic moment, $r$ represents the distance between the poles and $\vec{B}$ the cosmological magnetic field. \begin{eqnarray}
B(R)  = &\;\,B_0 \;\tanh{\left(\frac{R-R_C}{R_1}\right)} \exp{\left(-\left|\frac{R-R_C}{R_0}\right|\right)}  \hspace{1.5cm}&{\mbox  for} \; R > 0, 
\nonumber\\
  = &- B_0 \;\tanh{\left(\frac{R+R_C}{R_1}\right)} \exp{\left(-\left|\frac{R+R_C}{R_0}\right|\right)} \hspace{1.5cm}&{\mbox  for} \; R < 0 .  \nonumber \\
  & 
  \label{galaxyEq}
 \end{eqnarray}
In the left hand side of Fig. \ref{galaxy} the model for the  magnetic field of a galaxy described in Eq. \ref{galaxyEq} is shown~\cite{Vento:2020vsq} and in the right hand side  its gradient along the model galaxy.

\begin{figure}[htb]
\begin{center}
\includegraphics[scale= 0.95]{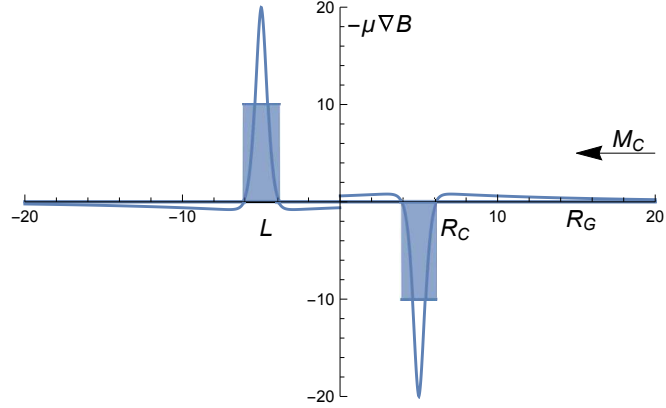} 
\end{center}
\caption{The simplified force field used in the calculation of the velocity is in grey over imposed on the real force field. $L$ is the width of the simplified force field.}
 \label{force}
\end{figure}

The mathematical structure of the field makes the analysis of the velocity of monopolium complicated. In order to get an estimate we substitute the  force field by two simplified forces of adequate strength as shown in Fig. \ref{force} compared to the actual force field. With this constant force field the velocity becomes,  after crossing the two accelerating structures, and using relativistic kinematics

\begin{equation}
v_n \sim \sqrt{1- \frac{1}{(\frac{g r_n B_0 L}{2 M_n R_1}+1)^2}} \sim \sqrt{1- \frac{1}{(0.32\, 10^{-26} \,r_n +1)^2}}\,.
\label{velocity}
\end{equation}
The last expression has been calculated for a high $n$ Rydberg state, $M_n \sim 2000$ GeV, and for conventional values of the galaxy parameters  $B_0 = 10 \mu$G $\sim1.95 \;10^{-25} $ GeV$^2$, $R_1= 0.5 $ Kpc, where the width of the force field $L \sim 2.2$ Kpc, with $r_n$ measured in fm.  Despite the huge sizes of Rydberg states ($r_n \sim 10^9$ fm $\sim 10^{-3}$ mm) the small pre-factor in the formula cannot be overcome and the velocity originating from the gravitational magnetic fields will be much smaller than the galactic gravitational velocity $v_M \sim 10^{-3} $. The latter  is mass independent~\cite{Preskill:1984gd}.

The  Rydberg states with long lifetimes and large magnetic moments are lightly bound monopole-antimonopole pairs with binding energies of the order of eVs and  the small Parker bound, an upper bound on the density of magnetic monopoles~\cite{Turner:1982ag}, can be obviated and detection is possible~\cite{Dicus:1983ry}. The Parker bound is based on the monopoles taking energy from the intergalactic magnetic field. However, monopolia below the dissociation energy do not take energy from the magnetic field since they are almost neutral, i.e., magnetic moment interactions are much weaker than magnetic charge interactions.

\begin{figure}[htb]
\begin{center}
\includegraphics[scale= 0.95]{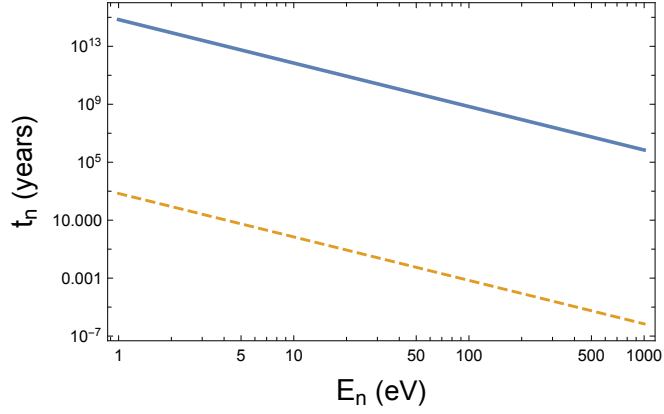} 
\end{center}
\caption{Lifetime of monopolium Rydberg states in years as a function of binding energies.}
\label{life}
\end{figure}

However, high lying Rydberg states might be dissociated by the much stronger Earth's magnetic field leading to free monopole-antimonopole pairs which might annihilate or ionize metal rods depending on its production kinematics. Monopolia below the Earth's dissociation limit survive and those close to the dissociation limit will have long lifetimes as shown in Fig. \ref{life} and we expect them to be able to transverse long coil systems.

Let us calculate the minimum binding energy before dissociation by equating  $E_n$ with the dissociation energy~\cite{Dicus:1983ry} 

\begin{equation}
E_n \sim g B_E r_n.
\label{dissociation}
\end{equation}
Using the $E_n = \frac{3 m}{4 \rho_n} = \frac{3 g^2}{4 r_n}$ and the Earth's magnetic field $B_E \sim 1$ G $\sim 1.95 \,10^{-20}$ GeV$^2$, we get for the maximum allowed size $r_n \lesssim \sqrt{\frac{3}{8 e B_E}} \sim 3.\, 10^{-3}$ mm.  For our light monopole $m \sim 10^3$ GeV we obtain for the binding energy $E_n  \gtrsim 1$ eV.

\begin{figure}[htb]
\begin{center}
\includegraphics[scale= 0.85]{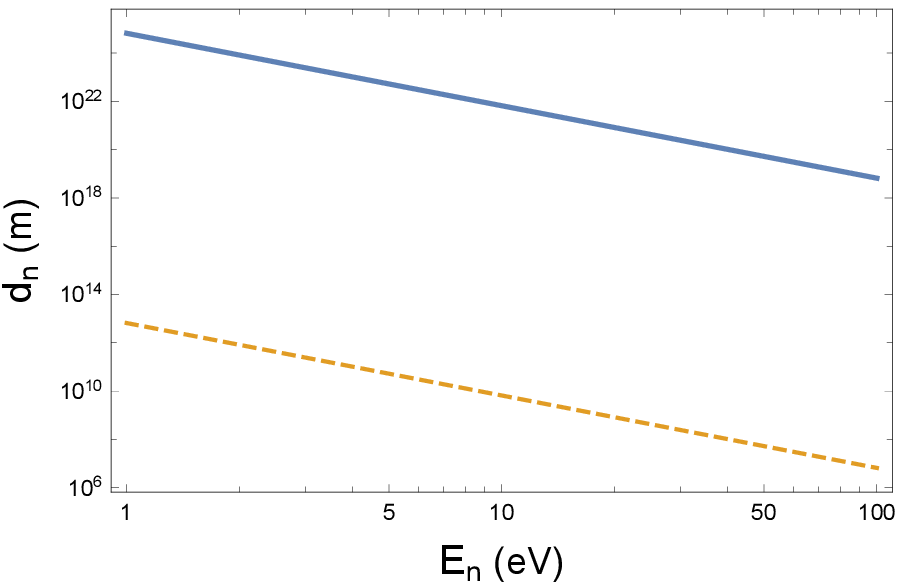} \hspace{0.5cm}\includegraphics[scale= 0.85]{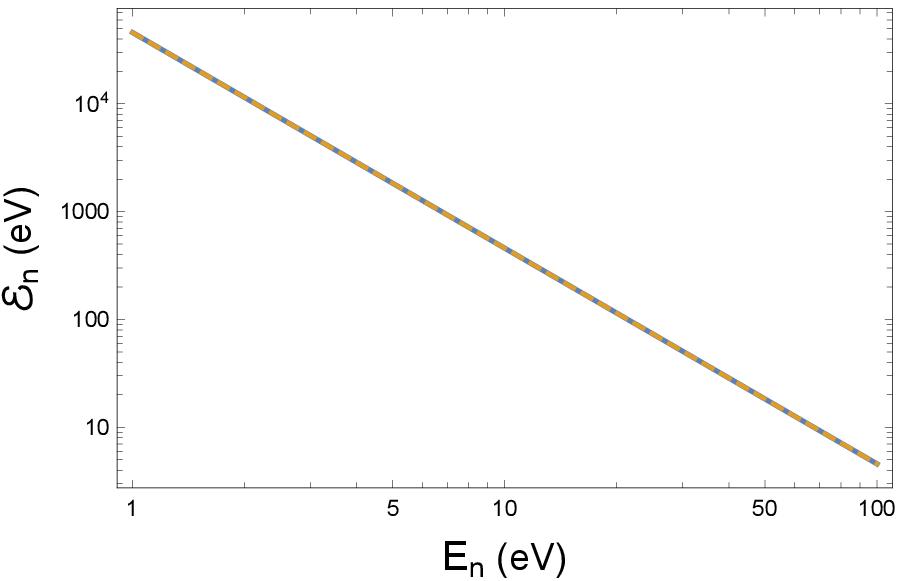} 
\end{center}
\caption{Left: We plot the distance traveled by monopolium before decaying as a function of binding energy. The solid curve represents the result in the Schwinger scheme and the dashed curve in the Dirac scheme. Right: The deposited energy in a $10^8$ coil system as a function of binding energy $E_n$. For small $E_n$ the deposited energy in the Schwinger (solid) and Dirac (dashed) schemes are equal.}
\label{CosmicMonopolium}
\end{figure}

Let us assume that we have one of those monopolia of $r_n \sim 10^{-3}$mm traversing at $v_M \sim 0.001$ a coil system of $40$ m long with $N \sim 10^8$ coils made of a good conductor $\varrho \sim 10^{-8}$ Ohm$\cdot$m, with $R\sim 0.01$ mm and $r_s \sim 0.0001$ mm.  In Fig. \ref{CosmicMonopolium} we show the distance traveled and most important the energy deposited in the coil system. For high Rydberg states close to the dissociation limit the deposited energy is large. This type of detector is suited in principle to detect single events in the line with the original monopole detectors \cite{Cabrera:1982gz,Milton:2006cp,MoEDAL:2014ttp,Acharya:2014nyr,Patrizii:2015uea}. Certainly a careful analysis of possible  backgrounds, a study which is outside the scope of the present investigation, has to be carried out.

\subsection{Cosmological heavy monopolium}

Let us pursue the study of cosmological  monopolium either heavy GUT or Kaluza Klein monopolium. The very high $n$  Rydberg states of monopolium below the  dissociation limit in the Earth's magnetic field have large lifetimes. The discussion related to the effect of the induced current on the particle velocity can be easily generalized also to heavy monopolium. The huge mass of the monopolium states leads to small accelerations and therefore the effect on the traveling particle velocity is here also negligible.

For GUT monopoles the Parker bound can be overcome when monopole-antimonopole bind. In this case, by equating the binding energy to the dissociation energy as above but using a mass for the monopole of $10^{16}$ GeV, we get  a limiting value for monopolia  of sizes $r_n  \lesssim10^{-3}$  mm which correspond to binding energies  $E_n \gtrsim 1$ eV as before. The velocity of the monopolia, in this case, is also the intergalactic gravitational  velocity $v_M \sim 10^{-3}$ which is independent of mass. The crucial ingredients of our previous calculation, binding energy and size,  remain the same and the analysis follows previous calculation for the very high Rydberg states leading to results similar to those shown in  Fig.\ref{CosmicMonopolium} for the deposited energy. The decay distance  is much larger since the ratio $(E_n/m)^3$ entering the spontaneous emission lifetime, Eq. (\ref{spontaneousemission}), is much smaller. Despite this difference the decay distances are so large for both light and GUT monopolia that their chances of detection are very similar and only the determination of their mass would distinguish them.

The idea behind Kaluza-Klein (KK) theories is that the world has more than three spatial dimensions and some of them are curled up to form a circle so small as to be unobservable  \cite{Kaluza:1921tu,Klein:1926tv}. KK theories have been the subject of revived interest in recent years since many standard model extensions in extra dimensions yield KK theories. KK theories contain a very rich topological structure, which includes very heavy monopoles whose mass is around the Planck mass  \cite{Gross:1983hb,Sorkin:1983ns}. Most important  they also contain other soliton solutions in different topological sectors. In particular  the dipole, which has the quantum numbers of a monopole-antimonopole bound state. As we have seen above, in conventional gauge theories monopolium has vacuum quantum numbers and annihilates. However, in KK theories monopolium does not belong to the topological sector of the vacuum and therefore it is classically stable  \cite{Gross:1983hb}.  In realistic scenarios the dipole does not have vacuum quantum numbers~\cite{Newman:1963yy,Hawking:1976jb,Gibbons:1979xm,Vento:2020vsq} and its structure is described in terms of some parameter $r$ (distance between center of the poles), which approaches the monopole-antimonopole structure as  $r$ becomes large. The dipole can thus be interpreted as a bound state with a mass smaller than twice the monopole mass. The parameter $r$ describes the magnetic moment of the dipole. The ground state (smallest mass) dipole, i.e. when $r\rightarrow 0$, has vanishingly small dipole moment and therefore it is electromagnetically neutral and thus has gravitational interaction only. When $r > 0$ we can ascribe these states to excited states of monopolium and in the large $r$ limit we are discussing, the equivalent of Rydberg states. Since those states are very much like an elongated  monopole-antimonopole structure separated by a distance $r$ the magnetic moment is 

\begin{equation}
{ \cal M}_r \sim g\, r.
\end{equation}

Once this similarity is exploited, again the parameters that play a role in the calculation are  the monopole mass $m \sim M_{Plack}\sim\sqrt{\frac{\hbar \,c}{G}} \sim 10^{19}$ GeV where $G$ is the gravitational constant  $1.32 \, 10^{-39}$ fm/GeV, the binding energy, $E_r = 2M - M_r$, and the mass independent gravitational velocity $v_r\sim 10^{-3}$. In this case again binding energies and sizes will be determined by avoiding dissociation.  Let us use for our calculation the mass formula for the dipole  of the Gross-Perry model \cite{Gross:1983hb,Vento:2020vsq}

\begin{equation}
M_r = 2 \left(m-\frac{M_r^2}{\sqrt{4 M_r^2 + \frac{r^2}{G^2}}}\right),
\end{equation}
where $m \sim \frac{M_{Planck}}{4 \sqrt{\alpha}}$ the monopole mass, with $M_{Planck} \sim 1.22 \, 10^{19}$ GeV, $M_r$ the monopolia mass and $r$ is given in fm and the masses in GeV. For large $r$ we obtain  the binding energy approximation for the Rydberg states

\begin{equation}
E_r \sim \frac{8 G m^2}{r}.
\end{equation}
Equating this energy to the dissociation energy $E_r = g B_E r \sim \frac{g B_E r}{0.197 GeV\cdot fm} $, where $B_E$ is measured in GeV$^2$ and $r$ in fm, we get the maximum monopolium size in the Earth at $ r = \sqrt{\frac{1.58 GeV \cdot fm G m^2}{g B_E}}$ which turns out to be $r \sim 10^{-2}$ mm with a binding energy is $E_r \sim 1 $ eV.  Size and binding energy are similar to the ones above. The physics is going to be the same, and therefore the experimental setup might be also suitable for this monopole dynamics. Only a kinematic detector will distinguish between the three types of monopoles studied.  The fact that the KK monopolia are stable does not make any difference for detection purposes because the lifetimes for all other scenarios is very large for Rydberg states. We foresee in this case two mechanisms of production given their huge mass, either a primordial mechanism which took place before inflation~\cite{Vento:2020vsq} or a cataclysmic collision of monopolium clouds after inflation. The former will produce mostly ground state or low excited monopolia with small values of $r$, very difficult to detect with the present setup. The  latter will produce highly excited states with large $r$, but very few in number. 
Again in this scenario we need detectors which look at extraordinary events and for monopolium a coil system might be adequate.

One question which might have arisen to the reader is if these large monopolium might be dissociated when passing through matter. Cosmic monopolium must traverse the atmosphere to reach any terrestrial detector. This question requires a deeper investigation but we give here some qualitative arguments. On the Earth's surface the distance between molecules is $10^{-6}$ mm, which means that large monopolia, $10^{-3}$ mm, interact with many molecules, whose size is order Angstrom $\sim 10^{-7}$ mm, thus they behave point like for Rydberg monopolia. The simplest model is that the atmosphere seen from a moving monopolium is an unpolarized gas of magnetic dipoles. Two types of interactions are of interest. One is the magnetic charge to magnetic moment interaction $\sim \frac{g \overrightarrow{m}\cdot \overrightarrow{r}}{r^3}$, where $\overrightarrow{m}$ is the magnetic moment of the molecule, and $\overrightarrow{r}$ the relative distance between the magnetic charge and the molecule. The other is the interaction of magnetic moments of monopolium $\overrightarrow{\mathcal M}$ with $\overrightarrow{m}$ of the molecule, which is a complex expression but which can be qualitatively approximated  by $ \frac{\overrightarrow{m}\cdot\overrightarrow{\mathcal M}}{r^3}$. We note that both of them depend on the magnetic moment of the molecules. If we consider the atmosphere as an unpolarized gas of molecules the two interactions will cancel on the average and therefore monopolium will remain bound and continue its motion unperturbed. A more detailed study will be presented elsewhere, but one can argue that for distances charge-molecule or molecule-monopolium of $10^{-7}$ mm the binding interaction $\frac{g^2}{r_M}$ is stronger and therefore dissociation does not occur, nor does monopolium slow down by binding itself with many molecules.

\section{Concluding remarks}
\label{conclusion}

Monopolium, a bound state of monopole-antimonopole, has no magnetic charge and in its ground state no magnetic moment, being therefore very difficult to detect directly. In conventional gauge theories it has the quantum numbers of the vacuum and annihilates into photons, and these disintegrations have been intensively studied \cite{Epele:2007ic,Epele:2008un,Epele:2012jn,Barrie:2016wxf,Fanchiotti:2017nkk,Barrie:2021rqa}. In Kaluza-Klein theories monopolium is classically stable and therefore very long lived \cite{Gross:1983hb,Vento:2020vsq} but its detection problems still persist since it only interacts gravitationally. 

 What happens with excited monopolium states? This paper deals with the study of the excited states of monopolium. Excited states have permanent magnetic moments and their interaction with magnetic and electric fields is known.  We have used a theoretical analysis of magnetic dipole detection to learn about excited monopolium states. Analyzing the detailed characteristics of excited monopoliun states we have discovered many properties, in particular for the high $n$ Rydberg states.
 
 The first part of the analysis deals with a theoretical description of the interaction of  magnetic moment with coils and solenoids both for conductors and superconductors.  Given that the difference between the two is not large and that our aim is to learn about the behavior of Rydberg states more than to perform a detailed experimental analysis of detection, we have continued only with conductors. 
 
Our analysis shows that only very high lying Rydberg states might be detectable with coil systems.  Those with a large lifetime, that could  escape the beam and be detected, given their small production cross section and the limited luminosity, cannot be produced at colliders \cite{Fanchiotti:2021xyj}. Therefore we have to resort to cosmic monopolium. 

Since bound states do not absorb energy from the galactic magnetic fields, monopolia can overcome the Parker bound. When these excited monopolia reach the Earth, the Earth's magnetic field might break them giving rise to monopole-antimonopole pairs, which might not annihilate since the excited states are very large, $10^{-3}-10^{-2}$ mm. Typical magnetic monopole detectors can be successful in seeing them~\cite{Cabrera:1982gz,Milton:2006cp,MoEDAL:2014ttp,Acharya:2014nyr,Patrizii:2015uea,Mitsou:2020hmt}. However, for larger binding energies the monopolium Rydberg states remain and they are ideal candidates for the setup presented here.  These states have large lifetimes and therefore appear stable for the detectors. We can use Faraday-Lenz coil systems to signal their presence once possible backgrounds have been eliminated. A natural property of excited states  is the emission of cascading photons. In our case since the Rydberg states have a very small binding energy the cascading photons will be of very small energy $ \sim  1-100 $eV. These photons would be a characteristic feature of excited monopolium in all three scenarios. Thus the Faraday-Lenz coil system should be supplemented by a photon detector~\cite{Mitsou:2020hmt}.

In concluding we have seen that excited states of monopolium have besides gravitational interaction an electromagnetic interaction associated with their magnetic moments. We have investigated here the properties of these states and have arrived to the conclusion that only the very high Rydberg states live long enough and have sufficiently large magnetic moment to be detected. 

\section*{Data availability}
All data generated or analysed during this study are included in this published article. 

\section*{Acknowledgement}
Vicente Vento would like to thank  Daniel Cano, Fernando Mart\'{\i}nez, Vassia Mitsou and Jos\' e Luis Tain, for useful conversations. HF and CAGC were partially supported by ANPCyT, Argentina. VV was supported in part  Ministerio de Ciencia e Innovaci\'on and Agencia Estatal de Investigaci\'on of Spain MCIN/AEI/10.13039/501100011033, European Regional Development Fund Grant No. PID2019-105439 GB-C21 and by GVA PROMETEO/2021/083 .


\begin{thebibliography}{10}

\bibitem{Dirac:1931qs}
P.~A.~M. Dirac, ``{Quantised singularities in the electromagnetic field,},''
  {\em Proc. R. Soc. Lond. A}, vol.~133, pp.~60--72, 1931.

\bibitem{Dirac:1948um}
P.~A.~M. Dirac, ``{The Theory of magnetic poles},'' {\em Phys. Rev.}, vol.~74,
  pp.~817--830, 1948.

\bibitem{tHooft:1974kcl}
G.~'t~Hooft, ``{Magnetic Monopoles in Unified Gauge Theories},'' {\em Nucl.
  Phys. B}, vol.~79, pp.~276--284, 1974.

\bibitem{Polyakov:1974ek}
A.~M. Polyakov, ``{Particle Spectrum in the Quantum Field Theory},'' {\em JETP
  Lett.}, vol.~20, pp.~194--195, 1974.

\bibitem{Georgi:1972cj}
H.~Georgi and S.~L. Glashow, ``{Unified weak and electromagnetic interactions
  without neutral currents},'' {\em Phys. Rev. Lett.}, vol.~28, p.~1494, 1972.

\bibitem{Preskill:1984gd}
J.~Preskill, ``{MAGNETIC MONOPOLES},'' {\em Ann. Rev. Nucl. Part. Sci.},
  vol.~34, pp.~461--530, 1984.

\bibitem{Drukier:1981fq}
A.~K. Drukier and S.~Nussinov, ``{Monopole Pair Creation in Energetic
  Collisions: Is It Possible?},'' {\em Phys. Rev. Lett.}, vol.~49, p.~102,
  1982.

\bibitem{Zeldovich:1978wj}
Y.~B. Zeldovich and M.~Y. Khlopov, ``{On the Concentration of Relic Magnetic
  Monopoles in the Universe},'' {\em Phys. Lett. B}, vol.~79, pp.~239--241,
  1978.

\bibitem{Vento:2007vy}
V.~Vento, ``{Hidden Dirac Monopoles},'' {\em Int. J. Mod. Phys. A}, vol.~23,
  pp.~4023--4037, 2008.
  
  \bibitem{Epele:2012jn}
L.~N. Epele, H.~Fanchiotti, C.~A.~G. Canal, V.~A. Mitsou, and V.~Vento,
  ``{Looking for magnetic monopoles at LHC with diphoton events},'' {\em Eur.
  Phys. J. Plus}, vol.~127, p.~60, 2012.

\bibitem{Kaluza:1921tu}
T.~Kaluza, ``{Zum Unit\"atsproblem der Physik},'' {\em Int. J. Mod. Phys. D},
  vol.~27, no.~14, p.~1870001, 2018.

\bibitem{Klein:1926tv}
O.~Klein, ``{Quantum Theory and Five-Dimensional Theory of Relativity. (In
  German and English)},'' {\em Z. Phys.}, vol.~37, pp.~895--906, 1926.

\bibitem{Gross:1983hb}
D.~J. Gross and M.~J. Perry, ``{Magnetic Monopoles in Kaluza-Klein Theories},''
  {\em Nucl. Phys. B}, vol.~226, pp.~29--48, 1983.

\bibitem{Vento:2020vsq}
V.~Vento, ``{Primordial monopolium as dark matter},'' {\em Eur. Phys. J. C},
  vol.~81, no.~3, p.~229, 2021.

\bibitem{Cabrera:1982gz}
B.~Cabrera, ``{First Results from a Superconductive Detector for Moving
  Magnetic Monopoles},'' {\em Phys. Rev. Lett.}, vol.~48, pp.~1378--1380, 1982.

\bibitem{Milton:2006cp}
K.~A. Milton, ``{Theoretical and experimental status of magnetic monopoles},''
  {\em Rept. Prog. Phys.}, vol.~69, pp.~1637--1712, 2006.

\bibitem{MoEDAL:2014ttp}
B.~Acharya {\em et~al.}, ``{The Physics Programme Of The MoEDAL Experiment At
  The LHC},'' {\em Int. J. Mod. Phys. A}, vol.~29, p.~1430050, 2014.

\bibitem{Acharya:2014nyr}
B.~Acharya {\em et~al.}, ``{The Physics Programme Of The MoEDAL Experiment At
  The LHC},'' {\em Int. J. Mod. Phys. A}, vol.~29, p.~1430050, 2014.

\bibitem{Patrizii:2015uea}
L.~Patrizii and M.~Spurio, ``{Status of Searches for Magnetic Monopoles},''
  {\em Ann. Rev. Nucl. Part. Sci.}, vol.~65, pp.~279--302, 2015.

\bibitem{Vento:2019auh}
V.~Vento and M.~Traini, ``{Scattering of charged particles off
  monopole\textendash{}anti-monopole pairs},'' {\em Eur. Phys. J. C}, vol.~80,
  no.~1, p.~62, 2020.

\bibitem{Apyan:2007wq}
A.~Apyan, A.~B. Apyan, and M.~Schmitt, ``{Detecting neutrino magnetic moments
  with conducting loops},'' {\em Phys. Rev. D}, vol.~77, p.~037901, 2008.

\bibitem{HIRAKAWA1973287}
H.~Hirakawa and A.~Futakawa, ``Inductance of superconducting solenoids,'' {\em
  Cryogenics}, vol.~13, no.~5, pp.~287--289, 1973.

\bibitem{Masterson:2017ab}
R.~E. Masterson, ``{Introduction to Nuclear Reactor Physics (1sr. ed.), CRC
  Press.},'' 2017.



\bibitem{Mavromatos:2020gwk}
N.~E. Mavromatos and V.~A. Mitsou, ``{Magnetic monopoles revisited: Models and
  searches at colliders and in the Cosmos},'' {\em Int. J. Mod. Phys. A},
  vol.~35, no.~23, p.~2030012, 2020.

\bibitem{Goebel1970}
C.~Goebel, {\em Quanta, In: Essays in Theoretical Physics, ed. by P.G.O.
  Freund, C.J. Goebel, Y. Nambu}.
\newblock Chicago: University of Chicago Press, 1970.

\bibitem{PhysRev.160.1257}
L.~I. Schiff, ``Quarks and magnetic poles,'' {\em Phys. Rev.}, vol.~160,
  pp.~1257--1262, Aug 1967.

\bibitem{Barrie:2016wxf}
N.~D. Barrie, A.~Sugamoto, and K.~Yamashita, ``{Construction of a model of
  monopolium and its search via multiphoton channels at LHC},'' {\em PTEP},
  vol.~2016, no.~11, p.~113B02, 2016.

\bibitem{Baines:2018ltl}
S.~Baines, N.~E. Mavromatos, V.~A. Mitsou, J.~L. Pinfold, and A.~Santra,
  ``{Monopole production via photon fusion and Drell\textendash{}Yan processes:
  MadGraph implementation and perturbativity via velocity-dependent coupling
  and magnetic moment as novel features},'' {\em Eur. Phys. J. C}, vol.~78,
  no.~11, p.~966, 2018.
\newblock [Erratum: Eur.Phys.J.C 79, 166 (2019)].

\bibitem{Barrie:2021rqa}
N.~D. Barrie, A.~Sugamoto, M.~Talia, and K.~Yamashita, ``{Searching for
  monopoles via monopolium multiphoton decays},'' {\em Nucl. Phys. B},
  vol.~972, p.~115564, 2021.

\bibitem{Epele:2007ic}
L.~N. Epele, H.~Fanchiotti, C.~A. Garcia~Canal, and V.~Vento, ``{Monopolium:
  The Key to monopoles},'' {\em Eur. Phys. J. C}, vol.~56, pp.~87--95, 2008.

\bibitem{Epele:2008un}
L.~N. Epele, H.~Fanchiotti, C.~A.~G. Canal, and V.~Vento, ``{Monopolium
  production from photon fusion at the Large Hadron Collider},'' {\em Eur.
  Phys. J. C}, vol.~62, pp.~587--592, 2009.

\bibitem{Dirac:1930bga}
P.~A.~M. Dirac, ``{On the Annihilation of Electrons and Protons},'' {\em Proc.
  Cambridge Phil. Soc.}, vol.~26, pp.~361--375, 1930.

\bibitem{Ore:1949te}
A.~Ore and J.~L. Powell, ``{Three photon annihilation of an electron - positron
  pair},'' {\em Phys. Rev.}, vol.~75, pp.~1696--1699, 1949.

\bibitem{Dicus:1983ry}
D.~A. Dicus and V.~L. Teplitz, ``{Circumvention of Parker's bound on galactic
  magnetic monopoles},'' {\em Nature}, vol.~303, pp.~408--409, 1983.

\bibitem{Schwinger:1976fr}
J.~S. Schwinger, K.~A. Milton, W.-y. Tsai, L.~L. DeRaad, Jr., and D.~C. Clark,
  ``{Nonrelativistic Dyon-Dyon Scattering},'' {\em Annals Phys.}, vol.~101,
  p.~451, 1976.

\bibitem{Fanchiotti:2017nkk}
H.~Fanchiotti, C.~A. Garc\'\i{}a~Canal, and V.~Vento, ``{Multiphoton
  annihilation of monopolium},'' {\em Int. J. Mod. Phys. A}, vol.~32, no.~35,
  p.~1750202, 2017.

\bibitem{Turner:1982ag}
M.~S. Turner, E.~N. Parker, and T.~J. Bogdan, ``{Magnetic Monopoles and the
  Survival of Galactic Magnetic Fields},'' {\em Phys. Rev. D}, vol.~26,
  p.~1296, 1982.

\bibitem{Sorkin:1983ns}
R.~d. Sorkin, ``{Kaluza-Klein Monopole},'' {\em Phys. Rev. Lett.}, vol.~51,
  pp.~87--90, 1983.

\bibitem{Newman:1963yy}
E.~Newman, L.~Tamburino, and T.~Unti, ``{Empty space generalization of the
  Schwarzschild metric},'' {\em J. Math. Phys.}, vol.~4, p.~915, 1963.

\bibitem{Hawking:1976jb}
S.~W. Hawking, ``{Gravitational Instantons},'' {\em Phys. Lett. A}, vol.~60,
  p.~81, 1977.

\bibitem{Gibbons:1979xm}
G.~W. Gibbons and S.~W. Hawking, ``{Classification of Gravitational Instanton
  Symmetries},'' {\em Commun. Math. Phys.}, vol.~66, pp.~291--310, 1979.


\bibitem{Fanchiotti:2021xyj}
H.~Fanchiotti, C.~A.~G.~Canal, M.~Traini and V.~Vento,
``{Looking for monopolium excited states with electromagnetic detectors},''
[arXiv:2110.02646 [hep-ph]].

\bibitem{Mitsou:2020hmt}
V.~A. Mitsou, ``{MoEDAL physics results and future plans},'' {\em PoS},
  vol.~CORFU2019, p.~009, 2020.

\end{thebibliography}
\end{document}